\documentclass[11pt,leqno]{article}

\usepackage{amsmath}
\usepackage{amsthm}
\usepackage{amssymb}
\usepackage[dvips]{graphicx,color}

\theoremstyle{plain}
\newtheorem{theorem}{Theorem}[section]
\newtheorem{lemma}[theorem]{Lemma}
\newtheorem{proposition}[theorem]{Proposition}
\newtheorem{corollary}[theorem]{Corollary}
\theoremstyle{definition}
\newtheorem{definition}[theorem]{Definition}
\theoremstyle{remark}
\newtheorem{remark}[theorem]{Remark}

\numberwithin{equation}{section}

\begin{document}

\title{\textbf{An operator-theoretical study of \\ the specific heat and the critical magnetic field in \\ the BCS-Bogoliubov model of superconductivity}}

\author{Shuji Watanabe\\
Division of Mathematical Sciences\\
Graduate School of Engineering, Gunma University\\
4-2 Aramaki-machi, Maebashi 371-8510, Japan\\
Email: shuwatanabe@gunma-u.ac.jp}

\date{}

\maketitle

\begin{abstract}
In the preceding paper, introducing a cutoff, the present author gave a proof of the statement that the transition to a superconducting state is a second-order phase transition in the BCS-Bogoliubov model of superconductivity on the basis of fixed-point theorems, and solved the long-standing problem of the second-order phase transition from the viewpoint of operator theory. In this paper we study the temperature dependence of the specific heat and the critical magnetic field in the model from the viewpoint of operator theory. We first show some properties of the solution to the BCS-Bogoliubov gap equation with respect to the temperature, and give the exact and explicit expression for the gap in the specific heat divided by the specific heat. We then show that it does not depend on superconductors and is a universal constant. Moreover, we show that the critical magnetic field is smooth with respect to the temperature, and point out the behavior of both the critical magnetic field and its derivative.

\medskip


\medskip

\end{abstract}

\section{Introduction and preliminaries}

In the physics literature, one differentiates the thermodynamic potential with respect to the temperature twice in order to show that the transition from a normal conducting state to a superconducting state is a second-order phase transition in the BCS-Bogoliubov model of superconductivity. Since the thermodynamic potential has the solution to the BCS-Bogoliubov gap equation in its form, one differentiates the solution with respect to the temperature twice without showing that the solution is differentiable with respect to the temperature. Therefore, if the solution were not differentiable with respect to the temperature, then one could not differentiate the solution with respect to the temperature, and hence one could not show that the transition is a second-order phase transition. This is why we need to show that the solution is differentiable with respect to the temperature twice as well as its existence and uniqueness.

Actually, as far as the present author knows, no one (except for the present author) showed that the solution is differentiable with respect to the temperature twice. Then, on the basis of fixed-point theorems, the present author \cite[Theorems 2.3 and 2.4]{watanabe-five} introduced a cutoff and showed that the solution is indeed partially differentiable with respect to the temperature twice, and gave an operator-theoretical proof of the statement that the transition from a normal conducting state to a superconducting state is a second-order phase transition. In this way, from the viewpoint of operator theory, the present author solved the long-standing problem of the second-order phase transition left unsolved for sixty-two years since the discovery of the BCS-Bogoliubov model.

In this paper we introduce a cutoff and study the temperature dependence both of the specific heat at constant volume and of the critical magnetic field in the BCS-Bogoliubov model of superconductivity from the viewpoint of operator theory. On the basis of fixed-point theorems, we first show some properties of the solution with respect to the absolute temperature $T$ both at sufficiently small $T$ and at $T$ in the neighborhood of the transition temperature $T_c$. We then give the exact and explicit expression for $\Delta C_V(T_c)/C_V^N(T_c)$. Here, $C_V^N(T_c)$ denotes the specific heat at constant volume at $T=T_c$, and  $\Delta C_V(T_c)$ its gap at $T=T_c$. We show  that $\Delta C_V(T_c)/C_V^N(T_c)$ does not depend on superconductors and is a universal constant in the BCS-Bogoliubov model. As far as the present author knows, one obtains the same results only when the potential $U(\cdot,\,\cdot)$ in \eqref{eqn:bcsgapeq} below is a constant in the physics literature. But we obtain the results even when the potential $U(\cdot,\,\cdot)$ is not a constant but a function. Moreover, we show that the critical magnetic field applied to type-I superconductors is of class $C^1$ both with respect to sufficiently small $T$ and with respect to $T$ in the neighborhood of the transition temperature $T_c$, and point out the behavior of the critical magnetic field and its derivative. We carry out their proofs on the basis of fixed-point theorems. As far as the present author knows, no one (except for the present author) showed that the critical magnetic field is differentiable with respect to $T$.

Here the BCS-Bogoliubov gap equation \cite{bcs, bogoliubov} is a nonlinear integral equation
\begin{equation}\label{eqn:bcsgapeq}
u(T,\,x)=\int_{\varepsilon}^{\hslash\omega_D}
\frac{U(x,\,\xi)\, u(T,\, \xi)}{\,\sqrt{\,\xi^2+u(T,\, \xi)^2\,}\,}\,
\tanh \frac{\,\sqrt{\,\xi^2+u(T,\, \xi)^2\,}\,}{2T}\, d\xi, \quad T \geq 0, \quad x \in [\varepsilon,\, \hslash\omega_D],
\end{equation}
where the solution $u$ is a function of the absolute temperature $T$ and the energy $x$, and $\omega_D$ stands for the Debye angular frequency and is a positive constant. The potential $U(\cdot,\,\cdot)$ satisfies $U(x,\,\xi)>0$ at all $(x,\,\xi) \in [\varepsilon, \, \hslash\omega_D]^2$. Throughout this paper we use the unit where the Boltzmann constant $k_B$ is equal to 1.

\begin{remark}
In \eqref{eqn:bcsgapeq}, we introduce a cutoff $\varepsilon>0$ and fix it. In the original BCS-Bogoliubov gap equation, one sets $\varepsilon=0$ and does not introduce the cutoff $\varepsilon>0$ since the effect of the region around the Fermi surface is very important in superconductivity (see, e.g., \cite{chen}). But, if we do not introduce the cutoff $\varepsilon>0$, then the first-order derivative of the thermodynamic potential with respect to $T$ diverges logarithmically only at  the transition temperature $T_c$, and hence the entropy also  diverges only at $T_c$. Therefore, the transition from a normal conducting state to a superconducting state at $T=T_c$ is not a second-order phase transition. This contradicts a lot of experimental results that the transition is a second-order phase transition without an external magnetic field. Therefore, we introduce the cutoff $\varepsilon>0$ and fix it. For more details, see Remarks \ref{rmk:varepsilon} and \ref{rmk:varepsilontwo} below.
\end{remark}

We consider the solution $u$ to the BCS-Bogoliubov gap equation as a function of $T$ and $x$, and deal with the integral with respect to the energy $\xi$ in \eqref{eqn:bcsgapeq}. Sometimes one considers the solution $u$ as a function of the absolute temperature and the wave vector, and accordingly deals with the integral with respect to the wave vector over the three dimensional Euclidean space $\mathbb{R}^3$. In this situation, the existence and uniqueness of the solution were established and studied in \cite{odeh, billardfano, vansevenant, bls, deugeihailoss, fhns, fhss, freijihaizlseiringer, hhss, hainzlloss, haizlseiringer, haizlseiringer2, haizlseiringer3}. For interdisciplinary reviews of the BCS-Bogoliubov model of superconductivity, see Kuzemsky \cite[Chapters 26 and 29]{kuzemsky3} and \cite {kuzemsky, kuzemsky2}. From the viewpoint operator theory, the present author studied the temperature dependence of the solution and showed the second-order phase transition in the BCS-Bogoliubov model of superconductivity (see \cite{watanabe-one, watanabe-two, watanabe-three, watanabe-four, watanabe-five}).

In this connection, the BCS-Bogoliubov gap equation plays a role similar to that of the Maskawa--Nakajima equation \cite{maskawa-nakajima-one, maskawa-nakajima-two}. If there is a nonnegative solution to the Maskawa--Nakajima equation (resp. to the BCS-Bogoliubov gap equation), then the massless abelian gluon model (resp. the BCS-Bogoliubov model) exhibits the spontaneous breaking of the chiral symmetry (resp. the $U(1)$ symmetry). If there is a unique solution $0$ to the Maskawa--Nakajima equation  (resp. to the BCS-Bogoliubov gap equation), then the massless abelian gluon model (resp. the BCS-Bogoliubov model) realizes the chiral symmetry (resp. the $U(1)$ symmetry). In fact, the Maskawa-Nakajima equation has attracted considerable interest in elementary particle physics, and is applied to many models such as a massless abelian gluon model, a massive abelian gluon model, a quantum chromodynamics (QCD)-like model, a technicolor model and a top quark condensation model. In Professor Maskawa's Nobel lecture, he stated the reason why he reconsidered the spontaneous chiral symmetry breaking in a renormalizable model of strong interaction. See the present author's paper \cite{watanabe-three} for an operator-theoretical treatment of the Maskawa-Nakajima equation. 

Let us deal with the BCS-Bogoliubov gap equation \eqref{eqn:bcsgapeq} with a constant potential $U(\cdot,\,\cdot)$. Let $U_1>0$ is a  positive constant and set $\displaystyle{ U(x,\,\xi)=U_1 }$ at all $(x,\,\xi) \in [\varepsilon, \, \hslash\omega_D]^2$. Then the BCS-Bogoliubov gap equation \eqref{eqn:bcsgapeq} is reduced to the simple gap equation \cite{bcs}
\begin{equation}\label{eqn:delta1}
1=U_1\int_{\varepsilon}^{\hslash\omega_D}
 \frac{1}{\,\sqrt{\,\xi^2+\Delta_1(T)^2\,}\,}\,
 \tanh \frac{\, \sqrt{\,\xi^2+\Delta_1(T)^2\,}\,}{2T}\,d\xi, \quad 0 \leq T \leq\tau_1,
\end{equation}
where the temperature $\tau_1>0$ is defined by (see \cite{bcs} and \cite{niwa, ziman})
\[
1=U_1\int_{\varepsilon}^{\hslash\omega_D}
\frac{1}{\,\xi\,}\,\tanh \frac{\xi}{\,2\tau_1\,}\,d\xi.
\]
Here the solution becomes a function of the temperature $T$ only, and so we denote the solution by $\Delta_1$.

Physicists and engineers studying superconductivity always assume that there is a unique nonnegative solution $\Delta_1$ to the simple gap equation \eqref{eqn:delta1} and that the solution $\Delta_1$ is of class $C^2$ with respect to $T$. And they differentiate the solution with respect to $T$ without showing that it is differentiable with respect to $T$. As far as the present author knows, no one except for the present author gave a mathematical proof for these assumptions; the present author \cite{watanabe-one, watanabe-five} applied the implicit function theorem to \eqref{eqn:delta1} and gave a mathematical proof:

\begin{proposition}[{\cite[Proposition 1.2]{watanabe-one}}]\label{prp:delta-one}
Let $U_1>0$ is a  positive constant and set $U(x,\,\xi)=U_1$ at all $(x,\,\xi) \in [\varepsilon, \, \hslash\omega_D]^2$. Then there is a unique nonnegative solution $\Delta_1: [\,0,\,\tau_1\,] \to [0,\,\infty)$ to the simple gap equation \eqref{eqn:delta1} such that the solution $\Delta_1$ is continuous and strictly decreasing with respect to the temperature $T$ on $[\,0,\,\tau_1\,]$. Moreover, the solution $\Delta_1$ is of class $C^2$ with respect to $T$ on $[\,0,\,\tau_1\,)$ and satisfies
\begin{eqnarray*}
& & \Delta_1(0) = \frac{\, \sqrt{\, \left(\hslash\omega_D-\varepsilon \, e^{1/U_1} \right)\left(\hslash\omega_D -\varepsilon \, e^{-1/U_1}\right) \, } \,}{\,\sinh\frac{1}{\,U_1\,}\,}, \quad \Delta_1(\tau_1)=0, \quad \Delta_1'(0)=\Delta_1''(0)=0,  \\
& & \lim_{T\uparrow \tau_1} \Delta_1'(T)=-\infty.
\end{eqnarray*}
\end{proposition}

\begin{remark}
We set $\Delta_1(T)=0$ at all $T>\tau_1$.
\end{remark}

We introduce another positive constant $U_2>0$. Let $0<U_1<U_2$ and set $U(x,\,\xi)=U_2$ at all $(x,\,\xi) \in [\varepsilon,\, \hslash\omega_D]^2$. Then a similar discussion implies that for $U_2$, there is a unique nonnegative solution $\Delta_2: [\,0,\,\tau_2\,] \to [0,\,\infty)$ to the simple gap equation
\begin{equation}\label{eqn:delta2}
1=U_2\int_{\varepsilon}^{\hslash\omega_D}
 \frac{1}{\,\sqrt{\,\xi^2+\Delta_2(T)^2\,}\,}\,
 \tanh \frac{\, \sqrt{\,\xi^2+\Delta_2(T)^2\,}\,}{2T}\,d\xi, \quad 0 \leq T \leq\tau_2.
\end{equation}
Here the temperature $\tau_2>0$ is defined by
\[
1=U_2\int_{\varepsilon}^{\hslash\omega_D}
\frac{1}{\,\xi\,}\,\tanh \frac{\xi}{\,2\tau_2\,}\,d\xi.
\]
Note that the solution $\Delta_2$ to \eqref{eqn:delta2} has properties similar to those of the solution $\Delta_1$ to \eqref{eqn:delta1}.

\begin{remark}
We also set $\Delta_2(T)=0$ at all $T>\tau_2$.
\end{remark}

\begin{lemma}[{\cite[Lemma 1.5]{watanabe-one}}] \    The inequality $\tau_1<\tau_2$ holds. If $0\leq T<\tau_2$, then $\Delta_1(T)<\Delta_2(T)$. If $T\geq \tau_2$, then $\Delta_1(T)=\Delta_2(T)=0$.
\end{lemma}

We next turn to the BCS-Bogoliubov gap equation \eqref{eqn:bcsgapeq}. We assume the following condition on the potential.
\begin{equation}\label{eqn:conditionU}
U(\cdot,\,\cdot) \in C([\varepsilon,\, \hslash\omega_D]^2), \qquad (0<) \; U_1 < U(x,\,\xi) < U_2 \quad \mbox{at all} \quad (x,\,\xi) \in [\varepsilon,\, \hslash\omega_D]^2.
\end{equation}
Fixing $T$ \   $(0 \leq T \leq \tau_2)$, we consider the Banach space $C[\varepsilon,\, \hslash\omega_D]$ consisting of continuous functions of the energy $x$ only, and deal with the following temperature dependent subset $V_T$:
\[
V_T=\left\{ u(T,\,\cdot) \in C[\varepsilon,\, \hslash\omega_D]: \; \Delta_1(T) \leq u(T,\,x) \leq \Delta_2(T) \; \mbox{at} \; x \in [\varepsilon,\, \hslash\omega_D] \right\}.
\]

\begin{remark}
The set $V_T$ depends on the temperature $T$.
\end{remark}

The present author gives another proof of the existence and uniqueness of the nonnegative solution to the BCS-Bogoliubov gap equation, and shows how the solution varies with the temperature.

\begin{theorem}[{\cite[Theorem 2.2]{watanabe-one}}] \label{thm:3-1}
Assume \eqref{eqn:conditionU} and let $T \in [0,\, \tau_2]$ be fixed. Then there is a unique nonnegative solution $u_0(T,\,\cdot) \in V_T$ to the BCS-Bogoliubov gap equation \eqref{eqn:bcsgapeq}:
\[
u_0(T,\, x)=\int_{\varepsilon}^{\hslash\omega_D}
\frac{U(x,\,\xi)\, u_0(T,\, \xi)}{\,\sqrt{\,\xi^2+u_0(T,\, \xi)^2\,}\,}\,
\tanh \frac{\,\sqrt{\,\xi^2+u_0(T,\, \xi)^2\,}\,}{2T}\, d\xi, \, \quad  x \in [\varepsilon,\, \hslash\omega_D].
\]
Consequently, the solution $u_0(T,\,\cdot)$ with $T$ fixed is continuous with respect to the energy $x$ and varies with the temperature as follows:
\[
\Delta_1(T) \leq u_0(T,\, x) \leq \Delta_2(T) \quad \mbox{at} \quad
(T,\,x) \in [0,\, \tau_2] \times [\varepsilon,\, \hslash\omega_D].
\]
\end{theorem}

The existence and uniqueness of the transition temperature $T_c$ were pointed out previously (see  \cite{fhns, hhss, haizlseiringer, vansevenant}). In our case, we can define it as follows.

\begin{definition}\label{dfn:tc}
Let $u_0(T,\,\cdot)$ be as in Theorem \ref{thm:3-1}. Then the transition temperature $T_c$ is defined by 
\[
T_c=\inf\{ T>0: \, u_0(T,\, x)=0 \quad \mbox{at all} \quad x \in [\varepsilon,\, \hslash\omega_D] \}.
\]
\end{definition}

Actually, Theorem \ref{thm:3-1} tells us nothing about continuity (or smoothness) of the solution $u_0$ with respect to the temperature $T$. From the viewpoint of operator theory, the present author \cite[Theorem 1.2]{watanabe-two} showed that $u_0$ is indeed continuous both with respect to $T$ and with respect to $x$ under the restriction that $T$ is sufficiently small. Moreover, under a similar restriction, the present author and Kuriyama \cite[Theorem 1.10]{watanabe-four} showed that the solution $u_0$ is partially differentiable with respect to $T$ twice, that the first-order and second-order partial derivatives of $u_0$ are both continuous with respect to $(T,\, x)$, and that $u_0$ is monotone decreasing with respect to $T$ from the viewpoint of operator theory. As mentioned before, the present author \cite[Theorems 2.3 and 2.4]{watanabe-five} showed that the solution is partially differentiable with respect to $T$ (in the neighborhood of the transition temperature $T_c$) twice, and gave a proof of the statement that the transition from a normal conducting state to a superconducting state is a second-order phase transition from the viewpoint of operator theory.

Let us turn to the thermodynamic potential. The thermodynamic potential $\Omega$ is given by the partition function $Z$: 
\[
\Omega=-T \ln Z.
\]
As mentioned before, we use the unit where the Boltzmann constant $k_B$ is equal to 1 throughout this paper. We fix both the chemical potential and the volume of our physical system, and so we consider the thermodynamic potential $\Omega$ as a function of the temperature $T$ only. Let $T_c$ be the transition temperature (see Definition \ref{dfn:tc}), and let $u_0$ be the solution to the BCS-Bogoliubov gap equation \eqref{eqn:bcsgapeq}. Then the thermodynamic potential $\Omega$ in the BCS-Bogoliubov model becomes
\[
\Omega(T)=\left\{ \begin{array}{ll}\displaystyle{
 \Omega_N(T)} \qquad &(T \geq T_c), \\
 \noalign{\vskip0.3cm} \displaystyle{
  \Omega_S(T)=\Omega_N(T)+\Psi(T)} \qquad &(0 \leq T \leq T_c),
\end{array}\right.
\]
where
\begin{eqnarray}\label{eq:omegan}
\Omega_N(T)
&=& -2N_0\int_{\displaystyle{\varepsilon}}^{\displaystyle{\hslash\omega_D}} \xi \,d\xi
-4N_0T\int_{\displaystyle{\varepsilon}}^{\displaystyle{\hslash\omega_D}}
 \ln\left( 1+e^{\displaystyle{-\xi/T}} \right)\, d\xi \\ \nonumber
& & +2\int_{\displaystyle{-\mu}}^{\displaystyle{-\hslash\omega_D}} \xi \, N(\xi)\,d\xi
 -2T\int_{\displaystyle{-\mu}}^{\displaystyle{-\hslash\omega_D}}
 N(\xi)\ln\left( 1+e^{\displaystyle{\, \xi/T}} \right)\, d\xi \\ \nonumber
& &-2T\int_{\displaystyle{\hslash\omega_D}}^{\infty} N(\xi)
 \ln\left( 1+e^{\displaystyle{-\xi/T}} \right)\, d\xi,\qquad T \geq 0,
\end{eqnarray}
and
\begin{eqnarray}\label{eqn:superthermopotential}
\Psi(T)
&=& -2N_0 \int_{\varepsilon}^{\hslash\omega_D} \left\{ \sqrt{\,\xi^2+u_0(T,\, \xi)^2\,}-\xi \right\} \, d\xi \\
& & +N_0 \int_{\varepsilon}^{\hslash\omega_D} \frac{u_0(T,\, \xi)^2}{\,\sqrt{\,\xi^2+u_0(T,\, \xi)^2\,}\,}\,
\tanh \frac{\,\sqrt{\,\xi^2+u_0(T,\, \xi)^2\,}\,}{2T}\, d\xi \nonumber \\
& & -4N_0 T \int_{\varepsilon}^{\hslash\omega_D} \ln
\frac{\, 1+e^{ -\sqrt{\,\xi^2+u_0(T,\, \xi)^2\,}/T } \,}{  1+e^{-\xi/T}  } \, d\xi, \quad 0 \leq T \leq T_c.
 \nonumber
\end{eqnarray}
Here, $\mu>0$ is the chemical potential and is a positive constant, $N(\xi)\geq 0$ stands for the density of states per unit energy at the energy $\xi$ \quad $(-\mu \leq \xi < \infty)$. We assume that $N(\xi)$ is constant at all $\xi \in [\varepsilon, \, \hslash\omega_D]$, and we set $N(\xi)=N_0$ at all $\xi \in [\varepsilon, \, \hslash\omega_D]$. Here, $N_0$ is a positive constant. Note that the function $\xi \mapsto N(\xi)$ is continuous on $[-\mu , \, \infty)$ and that $N(\xi)=O(\sqrt{\xi})$ as $\xi \to \infty$. So the integral $\displaystyle{ \int_{\displaystyle{\hslash\omega_D}}^{\infty} N(\xi) \ln\left( 1+e^{\displaystyle{-\xi/T}} \right)\, d\xi }$ above is well defined at $T>0$. Note that $\Psi(T_c)=0$ since $u_0(T_c \, ,\, x)=0$ at all $x \in [\varepsilon, \, \hslash\omega_D]$ (see Definition \ref{dfn:tc}).

\begin{remark}\label{rmk:gapcv}
If the solution $u_0$ to the BCS-Bogoliubov gap equation \eqref{eqn:bcsgapeq} is partially differentiable with respect to the temperature $T$ twice, then the thermodynamic potential $\Omega$ is differentiable with respect to $T$ twice. Then the specific heat at constant volume at $T$ is given by
\[
C_V(T)=-T \, \frac{ \, \partial^2 \Omega \,}{\partial T^2}(T).
\]
Therefore the gap $\Delta C_V$ in the specific heat at constant volume at the transition temperature $T_c$ is given by
\[
\Delta C_V = -T_c \,  \frac{ \, \partial^2 \Psi \,}{\partial T^2}(T_c).
\]
\end{remark}

\begin{remark}\label{rmk:varepsilon}
When we try to show the second-order phase transition, we need to differentiate the thermodynamic potential with respect to the temperature $T$ twice. The first-order derivative $(\partial \Psi/\partial T)$ of the thermodynamic potential $\Psi$ (see \eqref{eqn:superthermopotential}) with respect to $T$ has the following  term in it:
\begin{equation}\label{eq:term}
-N_0 \, \int_{\varepsilon}^{\hslash\omega_D} \frac{1}{\, \sqrt{\,\xi^2+u_0(T,\, \xi)^2\,} \,} \left\{ \frac{\partial}{\,\partial T \,} \, u_0(T,\, \xi)^2 \right\} \, d\xi.
\end{equation}
Note that $u_0(T_c\,,\, \xi)=0$ at all $\xi \in [\varepsilon, \, \hslash\omega_D]$ and that
\[
\left. \frac{\partial}{\,\partial T \,} \, u_0(T,\, \xi)^2 \right|_{T=T_c} = -v(\xi) < 0
\]
at all $\xi \in [\varepsilon, \, \hslash\omega_D]$. Here the function $v(\cdot)$ is that in Condition (C) of Section 2 and is continuous on $[\varepsilon, \, \hslash\omega_D]$. At $T=T_c$, the term \eqref{eq:term} therefore becomes
\[
-N_0 \, \int_{\varepsilon}^{\hslash\omega_D} \frac{1}{\, \sqrt{\,\xi^2+u_0(T_c ,\, \xi)^2\,} \,} \left[ \frac{\partial}{\,\partial T \,} \, u_0(T,\, \xi)^2 \right]_{T=T_c} \, d\xi =
N_0 \, \int_{\varepsilon}^{\hslash\omega_D} \frac{\, v(\xi)\,}{\, \xi \,} \, d\xi.
\]
If $\varepsilon=0$ and $v(\xi) =O\left( \xi^{\alpha} \right)$ as $\xi \downarrow 0$ with $\alpha>0$, then the integral just above converges. But, if $\varepsilon=0$, then the condition that $v(\xi) > 0$ at all $\xi \in [\varepsilon, \, \hslash\omega_D]=[0, \, \hslash\omega_D]$ (see Condition (C) of Section 2) implies that $v(0) > 0$, and hence this condition does not imply that $v(\xi) =O\left( \xi^{\alpha} \right)$ as $\xi \downarrow 0$ with $\alpha>0$. Therefore, if $\varepsilon=0$, then the integral just above diverges logarithmically since $v(\xi) > 0$ at all $\xi \in [0, \, \hslash\omega_D]$. This means that if $\varepsilon=0$, then the first-order derivative of the thermodynamic potential with respect to $T$ diverges logarithmically only at  the transition temperature $T_c$, and that the entropy also  diverges only at $T_c$. Therefore, if $\varepsilon=0$, then the transition from a normal conducting state to a superconducting state at $T=T_c$ is not a second-order phase transition. This contradicts a lot of experimental results that the transition is a second-order phase transition without an external magnetic field. This is why we introduce $\varepsilon>0$ in the thermodynamic potential. This means that the solution $u_0(T, \, x)$ to the BCS-Bogoliubov gap equation \eqref{eqn:bcsgapeq} is defined at $x \geq \varepsilon$, and hence that the range of integration in the right side of \eqref{eqn:bcsgapeq} is from $\varepsilon$ to $\hslash\omega_D$. This is why we introduce the cutoff $\varepsilon>0$ also in the BCS-Bogoliubov gap equation \eqref{eqn:bcsgapeq}.
\end{remark}

\begin{remark}\label{rmk:varepsilontwo}
If the potential $U$ is a constant, then the solution $u_0$ to the BCS-Bogoliubov gap equation \eqref{eqn:bcsgapeq} does not depend on the energy $\xi$ and depends on the temperature $T$ only  (see Proposition 1.1). This implies that the function $v(\cdot)$ above becomes a constant, i.e., $v(\xi)=v_0$. Here, $v_0$ is a positive constant. Then the term \eqref{eq:term} becomes
\begin{equation}\label{eqn:termconst}
-N_0 \,  \left\{ \frac{\partial}{\,\partial T \,} \, u_0(T)^2 \right\} \, \int_{\varepsilon}^{\hslash\omega_D} \frac{ d\xi  }{\, \sqrt{\,\xi^2+u_0(T)^2\,} \,}.
\end{equation}
Note that $u_0(T_c)=0$ and that
\[
\left. \frac{\partial}{\,\partial T \,} \, u_0(T)^2 \right|_{T=T_c} = -v_0 < 0.
\]
At $T=T_c$, the term \eqref{eqn:termconst} becomes
\begin{equation}\label{eq:diverge}
N_0  \, v_0 \, \int_{\varepsilon}^{\hslash\omega_D} \frac{\, d\xi \,}{\, \xi \,}.
\end{equation}
If $\varepsilon=0$, then the integral \eqref{eq:diverge} again diverges logarithmically. Therfore, if $\varepsilon=0$, then the first-order derivative of the thermodynamic potential with respect to $T$ again diverges logarithmically only at  the transition temperature $T_c$, and the transition from a normal conducting state to a superconducting state at $T=T_c$ is not a second-order phase transition. We again reach a contradiction. This is why we introduce the cutoff $\varepsilon>0$ both in the thermodynamic potential and in the BCS-Bogoliubov gap equation \eqref{eqn:bcsgapeq}.
\end{remark}

\section{Main results}

In the physics literature, one differentiates the solution to the BCS-Bogoliubov gap equation, the thermodynamic potential and the critical magnetic field with respect to the temperature without showing that they are differentiable with respect to the temperature. So we need to show that they are differentiable with respect to the temperature, as mentioned in the preceding section. 

We introduce the cutoff $\varepsilon>0$ and assume that the potential $U(\cdot,\,\cdot)$ satisfies \eqref{eqn:conditionU} throughout this paper. We denote by $z_0 >0$ a unique solution to the equation $\displaystyle{ \frac{2}{\,  z \,} = \tanh z }$
\quad  $(z>0)$. The value of $z_0$ is nearly equal to 2.07, and the inequality $\displaystyle{ \frac{2}{\,  z \,} \leq \tanh z }$ holds for $z \geq z_0$. Let  $\tau_0 \, (>0)$ satisfy
\begin{equation}\label{eq:tau0}
\Delta_1(\tau_0) = 2z_0\tau_0 \,.
\end{equation}
Let $0<\tau_3 < \tau_0$ and fix $\tau_3$. Here, $\tau_3>0$ is small enough. Let $\gamma$ be as in \eqref{eqn:gamma} below. We deal with the following subset $V$ of the Banach space $C([0,\, \tau_3] \times [\varepsilon,\,\hslash\omega_D])$:
\begin{eqnarray}
V &=& \left\{ u \in C([0,\, \tau_3] \times [\varepsilon,\,\hslash\omega_D]) :  0 \leq u(T,\,x)-u(T',\,x) \leq \gamma \left( T'-T \right) \; \; (T<T'),
\right.  \nonumber \\
& & \left. \Delta_1(T) \leq u(T,\,x) \leq \Delta_2(T), \   u \   \mbox{is partially differentiable with respect to} \  T \   \mbox{twice}, 
\right.  \nonumber \\
& & \left. \frac{\,\partial u\,}{\partial T}, \   \frac{\,\partial^2 u\,}{\partial T^2}  \in C([0,\, \tau_3] \times [\varepsilon,\,\hslash\omega_D]), \    \frac{\,\partial u\,}{\partial T}(0,\, x)=\frac{\,\partial^2 u\,}{\partial T^2}(0,\, x)=0 \    \mbox{at all } \    x \in [\varepsilon,\,\hslash\omega_D]
\right\}. \nonumber
\end{eqnarray}

\begin{remark}
The conditions in the definition of subset $V$
\[
\frac{\,\partial u\,}{\partial T}(0,\, x)=\frac{\,\partial^2 u\,}{\partial T^2}(0,\, x)=0 \    \mbox{at all } \    x \in [\varepsilon,\,\hslash\omega_D]
\]
are not imposed in \cite[Theorem 1.10]{watanabe-four}. These conditions are essential for our proof of Theorem \ref{thm:cmag} below. The other conditions in the definition of $V$ and in Theorem \ref{thm:main1} below are the same as the conditions in \cite[Theorem 1.10]{watanabe-four}.  
\end{remark}

We then define our operator $A$ (see \eqref{eqn:bcsgapeq}) on $V$:
\begin{equation}\label{eqn:ournlop}
Au(T,\,x)=\int_{\varepsilon}^{\hslash\omega_D}
\frac{U(x,\,\xi)\, u(T,\, \xi)}{\,\sqrt{\,\xi^2+u(T,\, \xi)^2\,}\,}\,
\tanh \frac{\,\sqrt{\,\xi^2+u(T,\, \xi)^2\,}\,}{2T}\, d\xi, \qquad u \in V.
\end{equation}
We denote by $\overline{V}$ the closure of the subset $V$ with respect to the norm of the Banach space $C([0,\, \tau_3] \times [\varepsilon,\,\hslash\omega_D])$.

The following is one of our main results.
\begin{theorem}\label{thm:main1}
Let us introduce the cutoff $\varepsilon>0$ and assume \eqref{eqn:conditionU}. Let $\overline{V}$ be as above. Then our operator $A:\, \overline{V} \to \overline{V}$ has a unique fixed point $u_0 \in \overline{V}$, and so there is a unique nonnegative solution $u_0 \in \overline{V}$ to the BCS-Bogoliubov gap equation \eqref{eqn:bcsgapeq}:
\[
u_0(T,\,x)=\int_{\varepsilon}^{\hslash\omega_D}
\frac{U(x,\,\xi)\, u_0(T,\, \xi)}{\,\sqrt{\,\xi^2+u_0(T,\, \xi)^2\,}\,}\,
\tanh \frac{\,\sqrt{\,\xi^2+u_0(T,\, \xi)^2\,}\,}{2T}\, d\xi, \quad (T,\, x) \in [0,\, \tau_3] \times [\varepsilon,\, \hslash\omega_D].
\]
Consequently, the solution $u_0$ is continuous on $[0,\, \tau_3] \times [\varepsilon,\, \hslash\omega_D]$. Moreover, $u_0$ is monotone decreasing and Lipschitz continuous with respect to $T$, and satisfies $\Delta_1(T) \leq u_0(T,\,x) \leq \Delta_2(T)$ at all $(T,\,x) \in [0,\, \tau_3] \times [\varepsilon,\,\hslash\omega_D]$. Furthermore, if $u_0 \in V$, then $u_0$ is partially differentiable with respect to $T$ twice, and the first-order and second-order partial derivatives of $u_0$ are both continuous on $[0,\, \tau_3] \times [\varepsilon,\, \hslash\omega_D]$. And, at all $x \in [\varepsilon,\,\hslash\omega_D]$,
\[
\frac{\,\partial u_0 \,}{\partial T}(0,\, x)=\frac{\,\partial^2 u_0 \,}{\partial T^2}(0,\, x)=0.
\]
On the other hand, if $u_0 \in \overline{V} \setminus V$, then $u_0$ is approximated by such a function of $V$ with respect to the norm of the Banach space $C([0,\, \tau_3] \times [\varepsilon,\,\hslash\omega_D])$.
\end{theorem}

\begin{remark}\label{rmk:uinvbar}
Let $u_0$ be the solution of Theorem \ref{thm:main1}. Since $u_0 \in \overline{V}$, we have $u_0 \in V$ or $u_0 \in \overline{V} \setminus V$. If  $u_0 \in V$, then the solution in the thermodynamic potential $\Psi(T)$ (see \eqref{eqn:superthermopotential}) is nothing but this $u_0 \in V$, and hence the solution in $\Psi(T)$ is partially differentiable with respect to the temperature $T$ twice. So we can differentiate the thermodynamic potential $\Psi(T)$ with respect to the temperature $T$ twice. On the other hand, if  $u_0 \in \overline{V} \setminus V$, then $u_0 \in \overline{V} \setminus V$ is approximated by a suitably chosen element $u_1 \in V$. In such a case, we replace the solution in $\Psi(T)$ by this element $u_1 \in V$. Let us remind here that the element $u_1 \in V$ is partially differentiable with respect to the temperature $T$ twice. Once we replace the solution in $\Psi(T)$ by this element $u_1 \in V$, we can again differentiate the thermodynamic potential $\Psi(T)$ with respect to the temperature $T$ twice. In this way, in both cases, we can differentiate the thermodynamic potential $\Psi(T)$, and hence $\Omega(T)$ with respect to the temperature $T$ twice. 
\end{remark}

\begin{remark}
The behavior of the solution $u_0$ given by Theorem \ref{thm:main1} is in good agreement with the experimental data.
\end{remark}

The function
\[
(T,\, x) \mapsto \int_{\varepsilon}^{\hslash\omega_D} \frac{U(x,\,\xi)}{\,\sqrt{\,\xi^2+\Delta_2(T)^2\,}\,}\,
 \tanh \frac{\, \sqrt{\,\xi^2+\Delta_2(T)^2\,}\,}{2T}\,d\xi
\]
is continuous, and it follows from \eqref{eqn:delta2} that
\[
\int_{\varepsilon}^{\hslash\omega_D} \frac{U(x,\,\xi)}{\,\sqrt{\,\xi^2+\Delta_2(T)^2\,}\,}\,
 \tanh \frac{\, \sqrt{\,\xi^2+\Delta_2(T)^2\,}\,}{2T}\,d\xi<1
\]
since $U(x,\,\xi) < U_2$ (see \eqref{eqn:conditionU}). Note that the function
\[
(T,\, x) \mapsto \frac{\, \Delta_2(\tau)^2 \,}{2\, \varepsilon^2} 
 \int_{\varepsilon}^{\hslash\omega_D} \frac{\, U(x,\,\xi) \,}{\xi}\, \tanh \frac{\xi}{\, 2T \,}\,  d\xi 
\]
is also continuous. Here, $0<\tau<T_c$. We then consider the sum of the two continuous functions above:
\[
\int_{\varepsilon}^{\hslash\omega_D} \frac{U(x,\,\xi)}{\,\sqrt{\,\xi^2+\Delta_2(T)^2\,}\,}\,
 \tanh \frac{\, \sqrt{\,\xi^2+\Delta_2(T)^2\,}\,}{2T}\,d\xi+
 \frac{\, \Delta_2(\tau)^2 \,}{2\, \varepsilon^2}  \int_{\varepsilon}^{\hslash\omega_D}
  \frac{\, U(x,\,\xi) \,}{\xi}\, \tanh \frac{\xi}{\, 2T \,}\,  d\xi.
\]
Note that the second term just above tends to zero as $\Delta_2(\tau)/\varepsilon$ goes to zero. Let $\tau$ be very close to $T_c$ and let $\Delta_2(\tau)/\varepsilon$ be very small so that the inequality
\[
\int_{\varepsilon}^{\hslash\omega_D} \frac{U(x,\,\xi)}{\,\sqrt{\,\xi^2+\Delta_2(T)^2\,}\,}\,
 \tanh \frac{\, \sqrt{\,\xi^2+\Delta_2(T)^2\,}\,}{2T}\,d\xi
 +\frac{\, \Delta_2(\tau)^2 \,}{2\, \varepsilon^2}  \int_{\varepsilon}^{\hslash\omega_D}
  \frac{\, U(x,\,\xi) \,}{\xi}\, \tanh \frac{\xi}{\, 2T \,}\,  d\xi<1
\]
holds true.

We then fix $\tau$ and $\varepsilon$, and we deal with the set $[\tau,\, T_c] \times [\varepsilon,\,\hslash\omega_D] \in \mathbb{R}^2$. Note that the left side of the inequality just above is a continuous function of $(T,\, x) \in [\tau,\, T_c] \times [\varepsilon,\,\hslash\omega_D]$. We set
\begin{eqnarray*}
\alpha &=& \max_{(T,\, x) \in [\tau,\, T_c] \times [\varepsilon,\,\hslash\omega_D] }
\left[ \int_{\varepsilon}^{\hslash\omega_D} \frac{U(x,\,\xi)}{\,\sqrt{\,\xi^2+\Delta_2(T)^2\,}\,}\,
 \tanh \frac{\, \sqrt{\,\xi^2+\Delta_2(T)^2\,}\,}{2T}\,  d\xi \right. \\
&  & \left. \qquad \qquad \qquad \qquad \qquad +\frac{\, \Delta_2(\tau)^2 \,}{2\, \varepsilon^2}  \int_{\varepsilon}^{\hslash\omega_D}
  \frac{\, U(x,\,\xi) \,}{\xi}\, \tanh \frac{\xi}{\, 2T \,}\,  d\xi \right].  \nonumber
\end{eqnarray*}
Therefore,
\begin{equation}\label{eq:alpha}
\alpha<1.
\end{equation}

\begin{remark}
We let $\tau$ be very close to $T_c$, and we let $\Delta_2(\tau)/\varepsilon$ be very small so that \eqref{eq:alpha} holds true.
\end{remark}

Let us consider the following condition.

\noindent \textbf{Condition (C).} \     Let $\tau$ and $\varepsilon$ be as above. An element $u \in C([\tau,\, T_c] \times [\varepsilon,\,\hslash\omega_D])$ is partially differentiable with respect to the temperature $T \in [\tau,\, T_c)$ twice, and the partial derivatives $(\partial u/\partial T)$ and $(\partial^2 u/\partial T^2)$ both belong to $C([\tau,\, T_c) \times [\varepsilon,\,\hslash\omega_D])$. Moreover, for the $u$ above, there are a unique $v \in C[\varepsilon,\,\hslash\omega_D]$ and a unique  $w \in C[\varepsilon,\,\hslash\omega_D]$ satisfying the following:

\noindent (C1) \   $v(x)>0$ at all $x \in [\varepsilon,\,\hslash\omega_D]$.\\
\noindent (C2) \   For an arbitrary $\varepsilon_1>0$, there is a $\delta>0$ such that $|T_c-T|<\delta$ implies
\[
\left| v(x)-\frac{\, u(T,\, x)^2 \,}{T_c-T} \right|<T_c\,\varepsilon_1 \,.
\]
Here, $\delta$ does not depend on $x \in [\varepsilon,\,\hslash\omega_D]$.\\
\noindent (C3) \   For an arbitrary $\varepsilon_1>0$, there is a $\delta>0$ such that $|T_c-T|<\delta$ implies
\[
\left| w(x)-\frac{\, -v(x)-\frac{\partial}{\,\partial T\,}\left\{u(T,\, x)^2\right\} \,}{T_c-T}  \right|<\varepsilon_1 \,.
\]
Here, $\delta$ does not depend on $x \in [\varepsilon,\,\hslash\omega_D]$. \\
\noindent (C4) \   For an arbitrarily large $R>0$, there is a $\delta>0$ such that $|T_c-T|<\delta$ implies
\[
-\frac{\, \partial u \,}{\partial T}(T,\, x) > R,
\]
Here, $\delta$ does not depend on $x \in [\varepsilon,\,\hslash\omega_D]$.

\begin{remark}\label{rmk:wutc0}
If $u \in C([\tau,\, T_c] \times [\varepsilon,\,\hslash\omega_D])$ satisfies Condition (C2), then $u(T_c \,,\, x)=0$ at all $x \in [\varepsilon,\,\hslash\omega_D]$.
\end{remark}

\begin{remark}
Condition (C2) implies the function $\displaystyle{ \frac{\, \partial \, u^2 \,}{\partial T}(T,\, x) }$ converges to $-v(x)$ $(<0)$ uniformly with respect to $x$ as $T \uparrow T_c$.
\end{remark}

\begin{remark}
Condition (C3) implies the function $\displaystyle{ \frac{\, \partial^2 \, u^2 \,}{\partial T^2}(T,\, x) }$ converges to $w(x)$ uniformly with respect to $x$ as $T \uparrow T_c$.
\end{remark}

\begin{remark}
Condition (C4) implies $\displaystyle{ \frac{\, \partial \, u \,}{\partial T}(T,\, x) \to -\infty }$  as $T \uparrow T_c$. Condition (C4) is not imposed in \cite[Theorem 2.3]{watanabe-five}. The other conditions in the definition of the subset $W$ below and in Theorem \ref{thm:solution} below are the same as the conditions in \cite[Theorem 2.3]{watanabe-five}.
\end{remark}

We denote by $W$ the following subset of the Banach space $C([\tau,\, T_c] \times [\varepsilon,\,\hslash\omega_D])$:
\begin{eqnarray}
W &=& \left\{ u \in C([\tau,\, T_c] \times [\varepsilon,\,\hslash\omega_D]) :  u(T,\,x) \geq u(T',\,x) \; \; (T<T'),
\right.  \nonumber \\
& & \left. \Delta_1(T) \leq u(T,\,x) \leq \Delta_2(T) \; \; \mbox{at} \; \; (T,\,x), \; (T',\,x) \in [\tau,\, T_c] \times [\varepsilon,\,\hslash\omega_D], \right. \nonumber \\
& & \left. u \; \mbox{satisfies Condition (C) above} \right\}, \nonumber
\end{eqnarray}
and we define our operator $A$ (see \eqref{eqn:bcsgapeq}) on $W$:
\begin{equation}\label{eqn:ouroperatorAW}
Au(T,\,x)=\int_{\varepsilon}^{\hslash\omega_D}
\frac{U(x,\,\xi)\, u(T,\, \xi)}{\,\sqrt{\,\xi^2+u(T,\, \xi)^2\,}\,}\,
\tanh \frac{\,\sqrt{\,\xi^2+u(T,\, \xi)^2\,}\,}{2T}\, d\xi, \qquad u \in W.
\end{equation}
We denote by $\overline{W}$ the closure of the subset $W$ with respect to the norm of the Banach space $C([\tau,\, T_c] \times [\varepsilon,\,\hslash\omega_D])$.

The following is one of our main results.

\begin{theorem}\label{thm:solution}
Let us introduce the cutoff $\varepsilon>0$ and assume \eqref{eqn:conditionU}. Let $\tau>0$ be very close to $T_c$ and let $\Delta_2(\tau)/\varepsilon>0$ be very small so that \eqref{eq:alpha} holds true. Then our operator $A: \overline{W} \to \overline{W}$ is a contraction operator. Consequently, there is a unique fixed point $u_0 \in \overline{W}$ of our operator $A: \overline{W} \to \overline{W}$, and so there is a unique nonnegative solution $u_0 \in \overline{W}$ to the BCS-Bogoliubov gap equation \eqref{eqn:bcsgapeq}:
\[
u_0(T,\,x)=\int_{\varepsilon}^{\hslash\omega_D}
\frac{U(x,\,\xi)\, u_0(T,\, \xi)}{\,\sqrt{\,\xi^2+u_0(T,\, \xi)^2\,}\,}\,
\tanh \frac{\,\sqrt{\,\xi^2+u_0(T,\, \xi)^2\,}\,}{2T}\, d\xi, \quad (T,\, x) \in [\tau, \, T_c] \times [\varepsilon, \, \hslash\omega_D] \,.
\]
The solution $u_0$ is continuous on $[\tau,\, T_c] \times [\varepsilon,\,\hslash\omega_D]$, and is monotone decreasing with respect to the temperature $T$. Moreover, $u_0$ satisfies that $\displaystyle{ \Delta_1(T) \leq u(T,\,x) \leq \Delta_2(T) }$ at all $(T,\, x) \in [\tau,\, T_c] \times [\varepsilon,\,\hslash\omega_D]$, and that $u_0(T_c \,,\, x)=0$ at all $x \in [\varepsilon,\,\hslash\omega_D]$. If $u_0 \in W$, then $u_0$ satisfies Condition (C). On the other hand, if $u_0 \in \overline{W} \setminus W$, then $u_0$ is approximated by such a function of $W$ with respect to the norm of the Banach space $C([\tau,\, T_c] \times [\varepsilon,\,\hslash\omega_D])$.
\end{theorem}

\begin{remark}\label{rmk:solution}
Let $u_0$ be the solution of Theorem \ref{thm:solution}. Suppose $u_0 \in W$. First, Condition (C2) implies $\displaystyle{ \frac{\, \partial \, u_0^2 \,}{\partial T}(T,\, x) }$ converges to $-v(x)$ $(<0)$ uniformly with respect to $x$ as $T \uparrow T_c$. Second, Condition (C3) implies $\displaystyle{ \frac{\, \partial^2 \, u_0^2 \,}{\partial T^2}(T,\, x) }$ converges to $w(x)$ uniformly with respect to $x$ as $T \uparrow T_c$. Finally, Condition (C4) implies $\displaystyle{ \frac{\, \partial \, u_0 \,}{\partial T}(T,\, x) \to -\infty }$  as $T \uparrow T_c$. Here both of $-v$ and $w$ depend on $u_0$. This behavior of the solution $u_0$ is in good agreement with the experimental data.
\end{remark}

\begin{remark}\label{rmk:uinwbar}
Let $u_0$ be the solution of Theorem \ref{thm:solution}. Since $u_0 \in \overline{W}$, we have $u_0 \in W$ or $u_0 \in \overline{W} \setminus W$. If  $u_0 \in W$, then the solution in the thermodynamic potential $\Psi(T)$ (see \eqref{eqn:superthermopotential}) is nothing but this $u_0 \in W$, and hence the solution in $\Psi(T)$ satisfies Condition (C). So we can differentiate the thermodynamic potential $\Psi(T)$ with respect to the temperature $T$ twice. On the other hand, if  $u_0 \in \overline{W} \setminus W$, then $u_0 \in \overline{W} \setminus W$ is approximated by a suitably chosen element $u_1 \in W$. In such a case, we replace the solution in $\Psi(T)$ by this element $u_1 \in W$. Let us remind here that the element $u_1 \in W$ satisfies Condition (C). Once we replace the solution in $\Psi(T)$ by this element $u_1 \in W$, we can again differentiate the thermodynamic potential $\Psi(T)$ with respect to the temperature $T$ twice. In this way, in both cases, we can differentiate the thermodynamic potential $\Psi(T)$, and hence $\Omega(T)$ with respect to the temperature $T$ twice. 
\end{remark}

Let $g: \, [0,\, \infty) \rightarrow \mathbb{R}$ be given by
\begin{equation}\label{eq:fng}
g(\eta)= \left\{ \begin{array}{ll}\displaystyle{
-\frac{1}{\,\eta^2\,}\left( \frac{\,\tanh\eta\,}{\eta}- \frac{1}{\,\cosh^2\eta \,} \right) }
  \qquad &(\eta>0),\\
 \noalign{\vskip0.3cm} \displaystyle{
-\frac{\,2\,}{\,3\,} } &(\eta=0).
\end{array}\right.
\end{equation}
Note that $g(\eta)<0$. As mentioned before, if the solution $u_0$ to the BCS-Bogoliubov gap equation \eqref{eqn:bcsgapeq} is partially differentiable with respect to the temperature $T$ twice, then the thermodynamic potential $\Omega$ is differentiable with respect to $T$ twice, and the specific heat at constant volume at $T$ is given by
\[
C_V(T)=-T \, \frac{ \, \partial^2 \Omega \,}{\partial T^2}(T).
\]
Therefore the gap $\Delta C_V$ in the specific heat at constant volume at the transition temperature $T_c$ is given by (see Remark \ref{rmk:gapcv})
\[
\Delta C_V = -T_c \,  \frac{ \, \partial^2 \Psi \,}{\partial T^2}(T_c).
\]

\begin{remark}
In the physics literature, one differentiates the thermodynamic potential to obtain the specific heat at constant volume without showing that the thermodynamic potential is differentiable with respect to $T$. Note that the thermodynamic potential has the solution to the BCS-Bogoliubov gap equation \eqref{eqn:bcsgapeq} in its form. In other words, one differentiates the thermodynamic potential with respect to $T$ without showing that the solution is differentiable with respect to $T$. But Combining Theorems \ref{thm:main1} and \ref{thm:solution} with Remarks \ref{rmk:uinvbar} and \ref{rmk:uinwbar}  implies that we can differentiate the solution $u_0$, and hence the thermodynamic potential $\Omega$ with respect to $T$ twice.
\end{remark}

\begin{theorem}\label{thm:dcv-over-cn}
Let $u_0$ be the solution to the BCS-Bogoliubov gap equation \eqref{eqn:bcsgapeq} given by Theorem \ref{thm:solution}. Let $\Delta C_V$ be the gap in the specific heat at constant volume at $T=T_c$, and let $C_V^N(T_c)$ be the  specific heat at constant volume at $T=T_c$ corresponding to normal conductivity, i.e., $C_V^N(T_c)=-T_c \, (\partial^2 \Omega_N / \partial T^2)(T_c)$. Then $\displaystyle{ \Delta C_V / C_V^N(T_c) }$ is explicitly and exactly given by the expression
\[
\frac{\Delta C_V}{ \, C_V^N(T_c) \, } = -\frac{N_0}{\, 32 \, (T_c)^2 \, J \,}
 \int_{\varepsilon/(2\, T_c)}^{\hslash\omega_D/(2\, T_c)} v(2T_c\,\eta)^2 \, g(\eta)\, d\eta \quad (>0),
\]
where
\begin{eqnarray*}
J &=& 2 \, \int_{\varepsilon/(2\, T_c)}^{\hslash\omega_D/(2\, T_c)} \frac{N_0 \, \eta^2}{\, \cosh^2 \eta \,} \, d\eta
 +\int_{-\mu/(2\, T_c)}^{-\hslash\omega_D/(2\, T_c)} \frac{\, N(2 \, T_c \,\eta) \, \eta^2 \,}{\, \cosh^2 \eta \,} \, d\eta \nonumber \\
& & +\int_{\hslash\omega_D/(2\, T_c)}^{\infty} \frac{\, N(2 \, T_c \,\eta) \, \eta^2 \,}{\, \cosh^2 \eta \,} \, d\eta, \nonumber
\end{eqnarray*}
and $v(\cdot)$ is that in Condition (C).
\end{theorem}

\begin{remark}
The function $v(\cdot)$ of Theorem \ref{thm:dcv-over-cn} corresponds to the solution $u_0$ to the BCS-Bogoliubov gap equation \eqref{eqn:bcsgapeq} given by Theorem \ref{thm:solution}.
\end{remark}

Theorem \ref{thm:dcv-over-cn} gives the explicit and exact expression for $\Delta C_V/ C_V^N(T_c)$. Note that the value $U(x,\,\xi)$ is nearly equal to a constant at all $(x,\,\xi) \in [\varepsilon,\, \hslash\omega_D]^2$ in some superconductors. Moreover, note that the value $\hslash\omega_D/(2T_c)$ is very large in many superconductors. The following then gives that the expression just above does not depend on superconductors and is a universal constant.

\begin{corollary}\label{crl:dcv-over-cn}
Assume $U(x,\,\xi)=U_0$ at all $(x,\,\xi) \in [\varepsilon,\, \hslash\omega_D]^2$, where $U_0>0$ is a constant. If $\hslash\omega_D/(2\, T_c) \simeq \infty$ and $\varepsilon/(2\, T_c) \simeq 0$, then
\[
\frac{\Delta C_V}{ \, C_V^N(T_c) \, } \simeq
 \frac{\,  \pi^2      \,}{\, 
 \displaystyle{ 4 \, \int_0^{\infty} \frac{\eta^2}{\, \cosh^2 \eta \,} \, d\eta } \; \displaystyle{ \int_0^{\infty} \{ -g(\eta) \} \, d\eta }  \,},
\]
which does not depend on superconductors and is a universal constant.
\end{corollary}

\begin{remark}
It is well known that $\Delta C_V/C_V^N(T_c) \simeq 12/\{ 7 \, \zeta(3) \}$ in the BCS-Bogoliubov model of superconductivity. Here, $s \mapsto \zeta(s)$ is the Riemann zeta function. Therefore, Corollary \ref{crl:dcv-over-cn} gives another expression for $\Delta C_V/C_V^N(T_c)$. Note that we use the unit where $k_B=1$. 
\end{remark}

Let us turn to the critical magnetic field applied to type-I superconductors. It is well known that superconductivity is destroyed even at a temperature $T$ less than the transition temperature $T_c$ when the sufficiently strong magnetic field is applied to type-I superconductors. It is also known that, at a fixed temperature $T$, superconductivity is destroyed when the applied magnetic field is stronger that the critical magnetic field $H_c(T)$, and that superconductivity is not destroyed when the magnetic field is weaker than $H_c(T)$. The critical magnetic field $H_c(\cdot)$ is a function of the temperature $T$, and $H_c(T) \geq 0$ at $T \leq T_c$. The critical magnetic field is related to $\Psi(T)$ (see \eqref{eqn:superthermopotential}) as follows:
\[
-\frac{1}{\, 8\pi \,}\, H_c(T)^2=\Psi(T), \quad \hbox{and hence} \quad H_c(T)=\sqrt{ \, -8\pi \Psi(T) \,}.
\]

\begin{remark}
In the physics literature, one differentiates the thermodynamic potential, and hence the critical magnetic field with respect to $T$ without showing that they are differentiable with respect to $T$. Note that the thermodynamic potential has the solution to the BCS-Bogoliubov gap equation \eqref{eqn:bcsgapeq} in its form. In other words, one differentiates the critical magnetic field with respect to $T$ without showing that the solution is differentiable with respect to $T$. But Combining Theorems \ref{thm:main1} and \ref{thm:solution} with Remarks \ref{rmk:uinvbar} and \ref{rmk:uinwbar}  implies that we can differentiate the solution $u_0$, and hence the critical magnetic field  $H_c(\cdot)$ with respect to $T$.
\end{remark}

The following gives the smoothness of the critical magnetic field with respect to $T$ and some of its properties.

\begin{theorem}\label{thm:cmag} 
Let $H_c(\cdot)$ be the critical magnetic field.

\noindent \rm{(A)} \    Let $u_0$ be the solution to the BCS-Bogoliubov gap equation \eqref{eqn:bcsgapeq} given by Theorem \ref{thm:solution}. Then the following \rm{(i)}, \rm{(ii)} and \rm{(iii)} hold true.

\rm{(i)} \hbox{} $H_c(\cdot) \in C^1[\tau, \, T_c]$. Consequently, $H_c(\cdot)$ is differentiable on $[\tau, \, T_c]$ with respect to the temperature $T$, and its first-order derivative is continuous on $[\tau, \, T_c]$. 

\rm{(ii)} \hbox{} $H_c(T_c)=0$, $\displaystyle{ \frac{\,\partial H_c \,}{\,\partial T \,}(T)<0 }$ at $T \in [\tau,\, T_c]$, and
\[
\frac{\,\partial H_c \,}{\,\partial T \,}(T_c)=
 -\sqrt{ \, -\frac{\pi N_0}{\, 2 \, T_c^2 \,}
 \int_{\varepsilon/(2\, T_c)}^{\hslash\omega_D/(2\, T_c)} v(2T_c\,\eta)^2 \, g(\eta)\, d\eta \,} \quad (< 0).
\]

\rm{(iii)} \hbox{} If $T \simeq T_c$ \hbox{} $(T \leq T_c)$, then
\[
H_c(T) \simeq  \left( 1-\frac{T}{\, T_c \,} \right) \sqrt{ -\frac{\, \pi N_0 \,}{2} 
 \int_{\varepsilon/(2\, T_c)}^{\hslash\omega_D/(2\, T_c)} v(2T_c\,\eta)^2 \, g(\eta)\, d\eta } \quad (\geq 0).
\]

\smallskip

\noindent \rm{(B)} \    Let $u_0$ be the solution to the BCS-Bogoliubov gap equation \eqref{eqn:bcsgapeq} given by Theorem \ref{thm:main1}. Then the following \rm{(iv)}, \rm{(v)} and \rm{(vi)} hold true.

\rm{(iv)} \hbox{} $H_c(\cdot) \in C^1[0,\, \tau_3]$. Consequently, $H_c(\cdot)$ is differentiable on $[0,\, \tau_3]$ with respect to the temperature $T$, and its first-order derivative is continuous on $[0,\, \tau_3]$. 

\rm{(v)} 
\[
H_c(0) = \sqrt{ \;  8\pi N_0 \int_{\varepsilon}^{\hslash\omega_D}
 \frac{\, \{ \, \sqrt{ \xi^2+u_0(0,\,\xi)^2 }-\xi \, \}^2  \,}{\, \sqrt{ \xi^2+u_0(0,\,\xi)^2 } \,}
 \, d\xi  \; }  \quad (>0).
\]

\rm{(vi)} \hbox{} $\displaystyle{ \frac{\,\partial H_c \,}{\,\partial T \,}(T)<0 }$ at $T \in (0,\,\tau_3]$, and
\[
\frac{\,\partial H_c \,}{\,\partial T \,}(0)=0.
\]
\end{theorem}

\begin{remark}
As far as the present author knows, no one has pointed out that $H_c(\cdot) \in C^1[\tau, \, T_c]$ and that $H_c(\cdot) \in C^1[0,\, \tau_3]$ exept for the present author. Moreover, as far as the present author knows, no one has given the exact and explicit expressions for $(\partial H_c/\partial T)(T_c)$ and 
for $H_c(0)$. Moreover, no one has shown that $(\partial H_c/\partial T)(0)=0$.  
\end{remark}

\begin{remark}
In the BCS-Bogoliubov model of superconductivity, one obtains
\[
H_c(T) \simeq 1.74 H_c(0) \left( 1-\frac{T}{\, T_c \,} \right)
\]
for $T \simeq T_c$ under the restriction that the potential $U(\cdot,\,\cdot)$ of the BCS-Bogoliubov gap equation \eqref{eqn:bcsgapeq} is a constant, i.e., under the restriction that the function $v(\cdot)$ of Part (iii) is a constant. But, without this restriction, Part (iii) of Theorem \ref{thm:cmag} gives another expression for the behavior of $H_c(T)$ as $T \simeq T_c$.
\end{remark}

The behavior of $H_c(\cdot)$ given by Theorem \ref{thm:cmag} is in good agreement with the experimental data. See Figure 1 for the behavior of $H_c(T)$.

\section{Proof of Theorem \ref{thm:main1}}

We prove Theorem \ref{thm:main1} in this section. Our proof is similar to that 
of \cite[Theorem 1.10]{watanabe-four}. We denote by $\| \cdot \|$ the norm of the Banach space $C([0,\, \tau_3] \times [\varepsilon,\,\hslash\omega_D])$. 

The function
\[
T \mapsto \int_{\varepsilon}^{\hslash\omega_D} \frac{1}{\,\sqrt{\,\xi^2+\Delta_1(T)^2\,}\,}\,
 \tanh \frac{\, \sqrt{\,\xi^2+\Delta_1(T)^2\,}\,}{2\tau_0}\,d\xi, \qquad T \in [0,\, \tau_3] 
\]
is continuous since the function $T \mapsto \Delta_1(T)$ is continuous. We then set
\begin{eqnarray}
a&=&\max_{0 \leq T \leq \tau_3} \int_{\varepsilon}^{\hslash\omega_D}
 \frac{1}{\,\sqrt{\,\xi^2+\Delta_1(T)^2\,}\,}\,
 \tanh \frac{\, \sqrt{\,\xi^2+\Delta_1(T)^2\,}\,}{2\tau_0}\,d\xi \quad (>0), \label{eq:a} \\
b&=&\frac{32\tau_3^2}{\, \Delta_1(\tau_3)^2 \,} \, \arctan \frac{\, \hslash\omega_D \,}{\Delta_1(\tau_3)} \quad (>0).
\nonumber
\end{eqnarray}
Hence, at all $T \in [0,\, \tau_3]$, 
\begin{eqnarray*}
1&=&U_1\int_{\varepsilon}^{\hslash\omega_D}
 \frac{1}{\,\sqrt{\,\xi^2+\Delta_1(T)^2\,}\,}\,
 \tanh \frac{\, \sqrt{\,\xi^2+\Delta_1(T)^2\,}\,}{2T}\,d\xi \nonumber \\
&>&U_1\int_{\varepsilon}^{\hslash\omega_D}
 \frac{1}{\,\sqrt{\,\xi^2+\Delta_1(T)^2\,}\,}\,
 \tanh \frac{\, \sqrt{\,\xi^2+\Delta_1(T)^2\,}\,}{2\tau_0}\,d\xi \nonumber \\
\end{eqnarray*}
since $T \leq \tau_3<\tau_0$. Therefore, $1>U_1a$. Here, $a$ is that in \eqref{eq:a}. Let us choose $U_2 \, (>U_1)$ such that $1>U_2a$ holds true. Set
\begin{equation}\label{eqn:gamma}
\gamma = \frac{\, U_2b\,}{\, 1-U_2a \,} \quad (>0).
\end{equation}

\begin{lemma}\label{lm:setvvar}
The subset $\overline{V}$ is bounded, closed, convex and nonempty.
\end{lemma}

\begin{proof}
We have only to show that the subset $\overline{V}$ is convex. Let $u, v \in \overline{V}$. Then there are sequences $\{ u_n \}, \, \{ v_n \} \subset V$ \    $(n=1,\,2,\,3,\,\cdots)$ satisfying $\| u-u_n \| \to 0$ and $\| v-v_n \| \to 0$ as $n \to \infty$.

\noindent \textit{Step 1}. \    We show $t \, u_n+(1-t)\, v_n \in V$ for $t \in [0,\, 1]$. It is easy to see that $t \, u_n+(1-t)\, v_n \in C([0,\, \tau_3] \times [\varepsilon,\,\hslash\omega_D])$. Since
\begin{eqnarray*}
 & &\{ t \, u_n(T,\,x)+(1-t)\, v_n(T,\,x) \}-\{ t \, u_n(T',\,x)+(1-t)\, v_n(T',\,x) \} \\
&=& t\, \{ u_n(T,\,x)-u_n(T',\,x) \}+(1-t)\,\{  v_n(T,\,x)-v_n(T',\,x) \} \quad (T<T'),
\end{eqnarray*}
it follows that
\[
0 \leq \{ t \, u_n(T,\,x)+(1-t)\, v_n(T,\,x) \}-\{ t \, u_n(T',\,x)+(1-t)\, v_n(T',\,x) \} \leq \gamma (T'-T).
\]
Obviously, $\Delta_1(T) \leq t \, u_n(T,\,x)+(1-t)\, v_n(T,\,x)  \leq \Delta_2(T)$. Moreover, $t \, u_n+(1-t)\, v_n$ is partially differentiable with respect to $T$ twice, and
\[
\frac{\partial}{\,\partial T\,} \left\{ t \, u_n+(1-t)\, v_n \right\}, \quad
\frac{\partial^2}{\,\partial T^2 \,} \left\{ t \, u_n+(1-t)\, v_n \right\} \in
C([0,\, \tau_3] \times [\varepsilon,\,\hslash\omega_D]).
\]
Furthermore, at all $x \in [\varepsilon,\,\hslash\omega_D]$,
\[
\frac{\partial \,\left\{ t \, u_n+(1-t)\, v_n \right\} \, }{\,\partial T\,}(0,\,x)=
t \,\frac{\,\partial u_n\,}{\,\partial T\,}(0,\,x)
+(1-t)\, \frac{\,\partial v_n \,}{\,\partial T\,}(0,\,x)=0,
\]
and
\[
\frac{\partial^2 \,\left\{ t \, u_n+(1-t)\, v_n \right\} \, }{\,\partial T^2 \,}(0,\,x)=
t \,\frac{\,\partial^2 u_n\,}{\,\partial T^2\,}(0,\,x)
+(1-t)\, \frac{\,\partial^2 v_n \,}{\,\partial T^2 \,}(0,\,x)=0.
\]
Thus \   $\displaystyle{ t \, u_n+(1-t)\, v_n \in V }$.

\noindent \textit{Step 2}. \    We next show $\displaystyle{ t \, u+(1-t)\, v \in \overline{V} }$. Since
\[
\| t \, u+(1-t)\, v-\left\{ t \, u_n+(1-t)\, v_n \right\} \| \leq
t\, \| u-u_n \|+(1-t)\, \| v-v_n \| \to 0 \quad (n \to \infty),
\]
it follows $\displaystyle{ t \, u+(1-t)\, v \in \overline{V} }$. Thus the subset $\overline{V}$ is convex. 
\end{proof}

A proof similar to that of \cite[Lemma 2.5]{watanabe-four} gives the following.
\begin{lemma}\label{lm:gamma}
Let  $(T, \, x), \, (T', \, x) \in [0,\, \tau_3] \times [\varepsilon,\,\hslash\omega_D]$, and let $T<T'$. If $u \in V$, then
\[
0 \leq Au(T,\,x)-Au(T',\,x) \leq \gamma \left( T'-T \right).
\]
\end{lemma}

A proof similar to that of \cite[Lemma 2.4]{watanabe-four} gives the following.
\begin{lemma}\label{lm:delta}
Let $u \in V$. Then $\Delta_1(T) \leq Au(T,\,x) \leq \Delta_2(T)$ at each $(T,\,x) \in [0,\, \tau_3] \times [\varepsilon,\,\hslash\omega_D]$.
\end{lemma}

A proof similar to that of \cite[Lemma 2.6]{watanabe-four} gives the following.
\begin{lemma}\label{lm:contin}
Let $u \in V$. Then $Au \in C([0,\, \tau_3] \times [\varepsilon,\,\hslash\omega_D])$.
\end{lemma}

A straightforward calculation gives the following.
\begin{lemma}
Let $u \in V$. Then $Au$ is partially differentiable with respect to $T$ twice $(0 \leq T \leq \tau_3)$, and
\[
\frac{\,\partial Au\,}{\partial T}, \   \frac{\,\partial^2 Au\,}{\partial T^2}  \in C([0,\, \tau_3] \times [\varepsilon,\,\hslash\omega_D]).
\]
\end{lemma}

\begin{lemma}\label{lm:first-derivative}
Let $u \in V$. Then, at all $x \in [\varepsilon,\,\hslash\omega_D]$,   
\[
\frac{\,\partial Au\,}{\partial T}(0,\, x)=\frac{\,\partial^2 Au\,}{\partial T^2}(0,\,x)=0.
\]
\end{lemma}

\begin{proof}
By the preceding lemma, $Au$ is partially differentiable with respect to $T$ twice.

\textit{Step 1.} \  We first show
\[
\frac{\,\partial Au\,}{\partial T}(0,\, x)=0.
\]
A straightforward calculation gives
\[
\frac{\,\partial Au\,}{\partial T}(T,\, x)=\int_{\varepsilon}^{\hslash\omega_D} U(x,\,\xi) \left( I_1+I_2+I_3 \right) \, d\xi,
\]
where
\begin{eqnarray*}
I_1 &=& \frac{\,\partial u\,}{\partial T}(T,\,\xi)  \, \frac{\xi^2}{\, \left\{ \, \xi^2+u(T,\, \xi)^2 \, \right\}^{3/2}\, }
 \tanh \frac{ \, \sqrt{ \, \xi^2+u(T,\, \xi)^2 \, } \,}{2T}, \nonumber \\
I_2 &=& \frac{\,\partial u\,}{\partial T}(T,\,\xi)  \, 
 \frac{u(T,\, \xi)^2}{\, 2T \left\{ \, \xi^2+u(T,\, \xi)^2 \, \right\} \cosh^2 \frac{ \, \sqrt{ \, \xi^2+u(T,\, \xi)^2 \, }  \, }{2T}  \,},  \nonumber \\
I_3 &=& -\frac{u(T,\, \xi)}{\, 2T^2 \cosh^2 \frac{ \, \sqrt{ \, \xi^2+u(T,\, \xi)^2 \, }  \, }{2T}  \,}. \nonumber
\end{eqnarray*}
At $T=0$,
\[
I_1=\frac{\,\partial u\,}{\partial T}(0,\,\xi)  \, \frac{\xi^2}{\, \left\{ \, \xi^2+u(0,\, \xi)^2 \, \right\}^{3/2}\, }
=0,
\]
since $\displaystyle{ \frac{\,\partial u\,}{\partial T}(0,\,\xi)=0 }$ at all $\xi \in [\varepsilon,\, \hslash\omega_D]$. The inequality $\displaystyle{ \frac{1}{\, \cosh z\,} \leq \frac{(2n)!}{\, z^{2n} \,} }$ \   $(n=1,\,2,\,3,\,\cdots)$ gives $I_2=I_3=0$ at $T=0$. Thus
\[
\frac{\,\partial Au\,}{\partial T}(0,\, x)=0.
\]

\textit{Step 2.} \  We next show
\[
\frac{\,\partial^2 Au\,}{\partial T^2}(0,\, x)=0.
\]
A straightforward calculation gives
\[
\frac{\,\partial^2 Au\,}{\partial T^2}(T,\, x)=\int_{\varepsilon}^{\hslash\omega_D} U(x,\,\xi) \left( J_1+J_2+J_3+J_4 \right) \, d\xi,
\]
where
\begin{eqnarray*}
J_1 &=& \frac{\xi^2}{\, \left( \, \xi^2+u^2 \, \right)^{3/2}\, } \left\{
 u_{TT}-3\frac{ \, u \, \left( u_T \right)^2 \, }{\xi^2+u^2} \right\}
 \tanh \frac{ \, \sqrt{ \, \xi^2+u^2 \, } \,}{2T}, \nonumber \\
J_2 &=& \frac{1}{ \, \left( \xi^2+u^2 \right) \cosh^2 \frac{\, \sqrt{\xi^2+u^2} \,}{2T} \, } \left( 
 \frac{ \, 2u \left( u_T \right)^2+u^2 \, u_{TT} \, }{2T}-\frac{ \, \xi^2 \, u_T+u^2 \, u_T \, }{2T^2}
\right), \nonumber \\
J_3 &=& \frac{1}{ \, \cosh^2 \frac{\, \sqrt{\xi^2+u^2} \,}{2T} \, } \left\{
 \frac{u \left( u_T \right)^2}{\, T\left( \xi^2+u^2 \right)^2 \,} \left( \frac{ \, \xi^2 \, }{2}-u^2 \right)
+\frac{u}{\, T^3 \, }-\frac{u_T}{\, 2T^2 \, }  \right\}, \nonumber \\
J_4 &=& \frac{ \, \tanh \frac{\, \sqrt{\xi^2+u^2} \,}{2T} \, }{ \, \cosh^2 \frac{\, \sqrt{\xi^2+u^2} \,}{2T} \, }
 \left\{
 \frac{u^2 \, u_T }{\, T^3 \sqrt{ \xi^2+u^2 } \,}
 -\frac{u^3 \, \left( u_T \right)^2 }{\, 2T^2 \left( \xi^2+u^2 \right)^{3/2} \,}
 -\frac{ \, u \, \sqrt{ \xi^2+u^2 } \, }{2T^4}
 \right\}. \nonumber
\end{eqnarray*}
Here, $u$ denotes $u(T,\,\xi)$, $u_T$ denotes $\displaystyle{ \frac{\partial u}{ \, \partial T \, }(T,\,\xi) }$ and $u_{TT}$ denotes $\displaystyle{ \frac{\partial^2 u}{ \, \partial T^2 \, }(T,\,\xi) }$.

Since $\displaystyle{ \frac{\,\partial u\,}{\partial T}(0,\,\xi)=\frac{\partial^2 u}{ \, \partial T^2 \, }(0,\,\xi)=0 }$ at all $\xi \in [\varepsilon,\, \hslash\omega_D]$, the inequality $\displaystyle{ \frac{1}{\, \cosh z\,} \leq \frac{(2n)!}{\, z^{2n} \,} }$ \   $(n=1,\,2,\,3,\,\cdots)$ gives $J_1=J_2=J_3=J_4=0$ at $T=0$. Thus
\[
\frac{\,\partial^2 Au\,}{\partial T^2}(0,\, x)=0.
\]
\end{proof}

We thus have the following.
\begin{lemma} \quad 
$\displaystyle{ AV \subset V}$.
\end{lemma}

A proof similar to that of \cite[Lemma 2.9]{watanabe-four} gives the following.
\begin{lemma}\label{lm:setaw}
The set $AV$ is relatively compact.
\end{lemma}

A proof similar to that of \cite[Lemma 2.10]{watanabe-four} gives the following.
\begin{lemma}\label{lm:acontinuous}
The operator $A:\,  V \to V$ is continuous.
\end{lemma}

We next extend the domain $V$ of our operator $A$ to its closure $\overline{V}$ with respect to the norm $\| \cdot \|$ of the Banach space $C([0,\, \tau_3] \times [\varepsilon,\,\hslash\omega_D])$. For $u \in \overline{V}$, there is a sequence $\{ u_n \}_{n=1}^{\infty} \subset V$ satisfying $\| u-u_n \| \to 0$ as $n \to \infty$. An argument similar to that in the proof of Lemma \ref{lm:acontinuous} gives $\{ Au_n \}_{n=1}^{\infty} \subset V$ is a Cauchy sequence. Hence there is an $Au \in \overline{V}$  satisfying $\| Au-Au_n \| \to 0$ as $n \to \infty$. Note that $Au$ does not depend on how to choose the sequence $\{ u_n \}_{n=1}^{\infty} \subset V$. We thus have the following.

\begin{lemma}
$A:\,  \overline{V} \to \overline{V}$.
\end{lemma}

A proof similar to that of \cite[Lemma 2.12]{watanabe-four} gives the following.
\begin{lemma}
For $u \in \overline{V}$,
\[
Au(T,\,x)=\int_{\varepsilon}^{\hslash\omega_D}
\frac{U(x,\,\xi)\, u(T,\, \xi)}{\,\sqrt{\,\xi^2+u(T,\, \xi)^2\,}\,}\,
\tanh \frac{\,\sqrt{\,\xi^2+u(T,\, \xi)^2\,}\,}{2T}\, d\xi.
\]
\end{lemma}

Lemmas \ref{lm:gamma}, \ref{lm:delta} and \ref{lm:contin} hold for each $u \in \overline{V}$ since  the set $\overline{V}$ is the closure of  $V$. 

\begin{lemma}
Let $u \in \overline{V}$. Then $Au \in C([0,\, \tau_3] \times [\varepsilon,\,\hslash\omega_D])$, and
\[
0 \leq Au(T,\,x)-Au(T',\,x) \leq \gamma \left( T'-T \right), \quad 0 \leq T < T' \leq \tau_3.
\]
Moreover, $\displaystyle{ \Delta_1(T) \leq Au(T,\,x) \leq \Delta_2(T) }$.
\end{lemma}

\begin{lemma}\label{lm:vvarequcon}
The set $A\overline{V}$ is uniformly bounded and equicontinuous, and hence the set $A\overline{V}$ is relatively compact.
\end{lemma}

\begin{proof}
Since $Au(T,\,x) \leq \Delta_2(0)$ for $u \in \overline{V}$, the set $A\overline{V}$ is uniformly bounded. By an argument similar to that in the proof of Lemma \ref{lm:contin}, the set $A\overline{V}$ is equicontinuous. Hence the set $A\overline{V}$ is relatively compact.
\end{proof}

By an argument similar to that in the proof of Lemma \ref{lm:acontinuous} gives the following.

\begin{lemma}\label{lm:avarcontin}
The operator $A:\,  \overline{V} \to \overline{V}$ is continuous.
\end{lemma}

Lemmas \ref{lm:vvarequcon} and \ref{lm:avarcontin} immediately imply the following.

\begin{lemma}\label{lm:acompact}
The operator $A:\,  \overline{V} \to \overline{V}$ is compact.
\end{lemma}

\begin{lemma}
The operator $A:\,  \overline{V} \to \overline{V}$ has a unique fixed point $u_0 \in \overline{V}$, i.e., $\displaystyle{ u_0=Au_0 }$.
\end{lemma}

\begin{proof}
Combing Lemma \ref{lm:acompact} with Lemma \ref{lm:setvvar} and applying the Schauder fixed-point theorem give that the operator $A:\,  \overline{V} \to \overline{V}$ has at least one fixed point $u_0 \in \overline{V}$. The uniqueness of $u_0 \in \overline{V}$ is pointed out in Theorem \ref{thm:3-1}. 
\end{proof}

Our proof of Theorem \ref{thm:main1} is now complete.

\section{Proof of Theorem \ref{thm:solution}}

We prove Theorem \ref{thm:solution} in this section. Our proof is similar to that of 
\cite[Theorem 2.3]{watanabe-five}. We denote by $\| \cdot \|$ the norm of the Banach space $C([\tau,\, T_c] \times [\varepsilon,\,\hslash\omega_D])$.

\smallskip

Let us show $A\, : \, W \to W$ first. A proof similar to that of \cite[Lemma 3.1]{watanabe-five} gives the following.
\begin{lemma}
If $u \in W$, then $Au \in C([\tau,\, T_c] \times [\varepsilon,\,\hslash\omega_D])$. 
\end{lemma}

A proof similar to that of Lemma \ref{lm:delta} gives the following.
\begin{lemma}
Let $(T,\, x) \in [\tau,\, T_c] \times [\varepsilon,\,\hslash\omega_D]$.
If $u \in W$, then $\Delta_1(T) \leq Au(T,\,x) \leq \Delta_2(T)$.
\end{lemma}

A proof similar to that of Lemma \ref{lm:gamma} gives the following.
\begin{lemma}
Let $(T,\,x), \, (T_1,\,x) \in [\tau,\, T_c] \times [\varepsilon,\,\hslash\omega_D]$, and let $T<T_1$.
If $u \in W$, then $Au(T,\,x) \geq Au(T_1,\,x)$.
\end{lemma}

In order to conclude $A\, : \, W \to W$, let us show that $Au$  satisfies Condition (C) for $u \in W$.
\begin{lemma}
Let $u \in W$. Then $Au$ is partially differentiable with respect to $T \in [\tau,\, T_c)$ twice, and
\[
\frac{\, \partial Au \,}{\partial T}, \;  \frac{\, \partial^2 Au \,}{\partial T^2} \in C([\tau,\, T_c) \times [\varepsilon,\,\hslash\omega_D]).
\]   
\end{lemma}

\begin{proof}
A straightforward calculation gives the result.
\end{proof}

Let $u \in W$ and let $v$ be as in Condition (C). Here, $v$ depends on the $u$. We set
\begin{equation}\label{eqn:functionF}
F(x)=\left\{ \int_{\varepsilon}^{\hslash\omega_D} U(x,\,\xi)
\frac{\, \sqrt{ v(\xi) }\, }{\xi} \, \tanh \frac{\xi}{\, 2T_c \, }\, d\xi \right\}^2 \quad (>0), \quad x \in [\varepsilon, \, \hslash\omega_D]. 
\end{equation}

A proof similar to that of \cite[Lemma 3.5]{watanabe-five} gives the following.
\begin{lemma}\label{lm:vandF}
Let $u \in W$, and let the function $F$ be as in \eqref{eqn:functionF}. Then the function $F$ belongs to $C[\varepsilon,\,\hslash\omega_D]$. Moreover, for an arbitrary $\varepsilon_1>0$, there is a $\delta>0$ such that $|T_c-T|<\delta$ implies
\[
\left| F(x)-\frac{\, \{ Au(T,\, x) \}^2 \,}{T_c-T} \right|<T_c\,\varepsilon_1 \,.
\]
Here, $\delta$ does not depend on $x \in [\varepsilon,\,\hslash\omega_D]$. Such a function $F$ is uniquely given by \eqref{eqn:functionF}.
\end{lemma}

Let $u \in W$. Let $v$ and $w$ be as in Condition (C), where both of $v$ and $w$ depend on the $u$. We set
\begin{eqnarray}\label{eqn:functionG}
& & \\
G(x) &=& \int_{\varepsilon}^{\hslash\omega_D} U(x,\,\xi)
 \frac{\, \sqrt{ v(\xi) }\, }{\xi} \, \tanh \frac{\xi}{\, 2T_c \, }\, d\xi \times \nonumber \\ \nonumber
  & & \hspace{-1cm} \int_{\varepsilon}^{\hslash\omega_D} U(x,\,\eta) \left\{
 \left( \frac{\, w(\eta)\, }{\,\eta\sqrt{\, v(\eta)\,}\,}-\frac{\, 2\sqrt{\,v(\eta)^3  \,} \,}{\eta^3} \right) 
 \, \tanh \frac{\eta}{\, 2T_c \, }+
 \frac{\sqrt{\, v(\eta)\,}}{\, \cosh^2\frac{\eta}{\, 2T_c\,} \,}\left(
 \frac{v(\eta)}{\, \eta^2T_c \,}+\frac{2}{\, T_c^2 \,} \right)
 \right\} \, d\eta,  \\ \nonumber
\end{eqnarray}
where $x \in [\varepsilon,\, \hslash\omega_D]$.

\begin{lemma}
Let $u \in W$, and let the function $G$ be as in \eqref{eqn:functionG}. Then the function $G$ belongs to $C[\varepsilon,\,\hslash\omega_D]$. Moreover, for an arbitrary $\varepsilon_1>0$, there is a $\delta>0$ such that $|T_c-T|<\delta$ implies
\[
\left| G(x)-\frac{\, -F(x)-\frac{\partial}{\,\partial T\,}\left\{ Au(T,\, x)\right\}^2 \,}{T_c-T}  \right|<\varepsilon_1 \,.
\]
Here, $\delta$ does not depend on $x \in [\varepsilon,\,\hslash\omega_D]$. Such a function $G$ is uniquely given by \eqref{eqn:functionG}.
\end{lemma}

\begin{proof}
The function $G$ belongs to $C[\varepsilon,\,\hslash\omega_D]$ since the potential $U(\cdot,\,\cdot)$ is uniformly continuous on $ [\varepsilon,\,\hslash\omega_D]^2$ by \eqref{eqn:conditionU}. An argument similar to that in the proof of Lemma \ref{lm:vandF} shows the rest. Here we also need Condition (C3).
\end{proof}

\begin{lemma}
Let $u \in W$. For an arbitrarily large $R>0$, there is a $\delta>0$ such that $|T_c-T|<\delta$ implies
\[
-\frac{\, \partial Au \,}{\partial T}(T,\, x) > R,
\]
Here, $\delta$ does not depend on $x \in [\varepsilon,\,\hslash\omega_D]$.
\end{lemma}

\begin{proof}
Let $u \in W$. An argument similar to that in the proof of Lemma \ref{lm:first-derivative} gives
\[
-\frac{\,\partial Au\,}{\partial T}(T,\, x)=-\int_{\varepsilon}^{\hslash\omega_D} U(x,\,\xi) \left( I_1+I_2+I_3 \right) \, d\xi,
\]
where
\begin{eqnarray*}
I_1 &=& \frac{\,\partial u\,}{\partial T}(T,\,\xi)  \, \frac{\xi^2}{\, \left\{ \, \xi^2+u(T,\, \xi)^2 \, \right\}^{3/2}\, }
 \tanh \frac{ \, \sqrt{ \, \xi^2+u(T,\, \xi)^2 \, } \,}{2T}, \nonumber \\
I_2 &=& \frac{\,\partial u\,}{\partial T}(T,\,\xi)  \, 
 \frac{u(T,\, \xi)^2}{\, 2T \left\{ \, \xi^2+u(T,\, \xi)^2 \, \right\} \cosh^2 \frac{ \, \sqrt{ \, \xi^2+u(T,\, \xi)^2 \, }  \, }{2T}  \,},  \nonumber \\
I_3 &=& -\frac{u(T,\, \xi)}{\, 2T^2 \cosh^2 \frac{ \, \sqrt{ \, \xi^2+u(T,\, \xi)^2 \, }  \, }{2T}  \,}. \nonumber
\end{eqnarray*}
Since $u$ satisfies Condition (C4), for an arbitrarily large $R>0$, there is a $\delta_1>0$ such that $|T_c-T|<\delta_1$ implies
\[
-\frac{\, \partial u \,}{\partial T}(T,\, x) > R,
\]
Here, $\delta_1$ does not depend on $x \in [\varepsilon,\,\hslash\omega_D]$. Then $I_1,\, I_2, \, I_3<0$,  and hence
\[
-\frac{\,\partial Au\,}{\partial T}(T,\, x) \geq -\int_{\varepsilon}^{\hslash\omega_D} U(x,\,\xi) \, I_1 \, d\xi
\quad (>0).
\]
Note that the function $\displaystyle{ z \mapsto \frac{\,\tanh z\,}{z} }$ \   $(z \geq 0)$ is strictly decreasing. Therefore,
\begin{eqnarray*}
& &-\int_{\varepsilon}^{\hslash\omega_D} U(x,\,\xi) \, I_1 \, d\xi \nonumber \\
&=& \int_{\varepsilon}^{\hslash\omega_D} U(x,\,\xi)
 \left\{ -\frac{\,\partial u\,}{\partial T}(T,\,\xi) \right\} \, \frac{\xi^2}{\, \left\{ \, \xi^2+u(T,\, \xi)^2 \,       
 \right\}^{3/2}\, } \tanh \frac{ \, \sqrt{ \, \xi^2+u(T,\, \xi)^2 \, } \,}{2T} \, d\xi \nonumber \\
&\geq& R \, U_1 \, \frac{\tau}{\, T_c \, \sqrt{ \, \hslash^2\omega_D^2+\Delta_2(0)^2 \, } \, }
 \tanh \frac{ \, \sqrt{ \, \hslash^2\omega_D^2+\Delta_2(0)^2 \, } \,}{2\tau} \,
 \int_{\varepsilon}^{\hslash\omega_D} \frac{\xi^2}{\, \xi^2+\Delta_2(0)^2 \, } \, d\xi.
\end{eqnarray*}
Since $R>0$ is arbitrarily large, the result follows. 
\end{proof}

The lemmas above immediately give the following.
\begin{lemma}
$\displaystyle{\quad A\, : \, W \to W. }$
\end{lemma}

We denote by $\| \cdot \|$ the norm of the Banach space $C([\tau,\, T_c] \times [\varepsilon,\,\hslash\omega_D])$, as mentioned above. A proof similar to that of \cite[Lemma 3.8]{watanabe-five} gives the following.
\begin{lemma}\label{lm:estimate}
Let $\alpha$ be as in \eqref{eq:alpha}, and let $u,\,v \in W$. Then \    $\left\| Au-Av \right\| \leq \alpha \| u-v \|$.
\end{lemma}

We extend the domain $W$ of our operator $A$ to its closure $ \overline{W}$. Let $u \in \overline{W}$. Then there is a sequence $\{ u_n \}_{n=1}^{\infty} \subset W$ satisfying $\| u-u_n \| \to 0$ as $n \to \infty$. By Lemma \ref{lm:estimate}, the sequence $\{ Au_n \}_{n=1}^{\infty} \subset W$ becomes a Cauchy sequence, and hence there is an $Au \in \overline{W}$  satisfying $\| Au-Au_n \| \to 0$ as $n \to \infty$. A straightforward calculation gives that $Au$ does not depend on the sequence $\{ u_n \}_{n=1}^{\infty} \subset W$. Thus we have the following.

\begin{lemma}
$A:\,  \overline{W} \to \overline{W}$.
\end{lemma}

A proof similar to that of \cite[Lemma 3.10]{watanabe-five} gives the following.
\begin{lemma}\label{lm:integrali}
Let $u \in \overline{W}$. Then
\[
Au(T,\,x)=\int_{\varepsilon}^{\hslash\omega_D}
\frac{U(x,\,\xi)\, u(T,\, \xi)}{\,\sqrt{\,\xi^2+u(T,\, \xi)^2\,}\,}\,
\tanh \frac{\,\sqrt{\,\xi^2+u(T,\, \xi)^2\,}\,}{2T}\, d\xi.
\]
\end{lemma}

From Lemma \ref{lm:estimate}, we immediately have the following.

\begin{lemma}\label{lm:estimate-prime}
Let $\alpha$ be as in \eqref{eq:alpha}, and let $u,\,v \in \overline{W}$. Then \    $\left\| Au-Av \right\| \leq a \| u-v \|$. Consequently, our operator $A: \, \overline{W} \to \overline{W}$ is a contraction operator.
\end{lemma}

Since our operator $A: \, \overline{W} \to \overline{W}$ is a contraction operator, the Banach fixed-point theorem thus implies the following.

\begin{lemma}
The operator $A: \overline{W} \to \overline{W}$ has a unique fixed point $u_0 \in \overline{W}$. 
Consequently, there is a unique nonnegative solution $u_0 \in \overline{W}$ to the BCS-Bogoliubov gap equation \eqref{eqn:bcsgapeq}:
\[
u_0(T,\,x)=\int_{\varepsilon}^{\hslash\omega_D}
\frac{U(x,\,\xi)\, u_0(T,\, \xi)}{\,\sqrt{\,\xi^2+u_0(T,\, \xi)^2\,}\,}\,
\tanh \frac{\,\sqrt{\,\xi^2+u_0(T,\, \xi)^2\,}\,}{2T}\, d\xi, \quad (T,\, x) \in [\tau, \, T_c] \times [\varepsilon, \, \hslash\omega_D].
\]
\end{lemma}

Now our proof of Theorem \ref{thm:solution} is complete.

\section{Proofs of Theorem \ref{thm:dcv-over-cn} and Corollary \ref{crl:dcv-over-cn}}

\quad \textit{Proof of Theorem \ref{thm:dcv-over-cn} } \    We first give a proof of Theorem \ref{thm:dcv-over-cn}. The thermodynamic potential $\Omega_N$ corresponding to normal conductivity is given by \eqref{eq:omegan}. The specific heat at constant volume at the temperature $T$ is defined by $\displaystyle{C_V(T)=-T \, \frac{ \, \partial^2 \Omega \,}{\partial T^2}(T) }$ (see Remark \ref{rmk:gapcv}). Then the specific heat at constant volume corresponding to normal conductivity is given by
\[
C_V^N(T)=-T \, \frac{ \, \partial^2 \Omega_N \,}{\partial T^2}(T), \qquad T \geq T_c \,.
\]

\begin{lemma} \    Let $\Omega_N$ be as in \eqref{eq:omegan}. Then
\begin{eqnarray*}
\frac{ \, \partial \Omega_N \,}{\partial T}(T)
&=& -4N_0 \int_{\varepsilon}^{\hslash\omega_D} \left\{ 
 \frac{\xi}{\, T\,}\left( -1+\frac{1}{\, e^{\xi/T}+1 \,} \right)+\ln \left(  e^{\xi/T}+1 \right) \right\} \, d\xi \\
& & -2 \int_{-\mu}^{-\hslash\omega_D} N(\xi) \left\{ 
 \frac{\xi}{\, T\,}\left( 1-\frac{1}{\, e^{-\xi/T}+1 \,} \right)+\ln \left(  e^{-\xi/T}+1 \right) \right\} \, d\xi \\
& & -2 \int_{\hslash\omega_D}^{\infty} N(\xi) \left\{ 
 \frac{\xi}{\, T\,}\left( -1+\frac{1}{\, e^{\xi/T}+1 \,} \right)+\ln \left(  e^{\xi/T}+1 \right) \right\} \, d\xi.
\end{eqnarray*}
Moreover,
\begin{eqnarray*}
\frac{ \, \partial^2 \Omega_N \,}{\partial T^2}(T)
&=& - \frac{N_0}{\, T^3 \,} \int_{\varepsilon}^{\hslash\omega_D}
 \frac{\xi^2}{\, \cosh^2 \left( \,\xi/(2T) \,\right) \,} \, d\xi
 - \frac{1}{\, 2T^3 \,} \int_{-\mu}^{-\hslash\omega_D}
 \frac{N(\xi) \, \xi^2}{\, \cosh^2 \left( \,\xi/(2T) \,\right) \,} \, d\xi \\
& & - \frac{1}{\, 2T^3 \,} \int_{\hslash\omega_D}^{\infty}
 \frac{N(\xi) \, \xi^2}{\, \cosh^2 \left( \,\xi/(2T) \,\right) \,} \, d\xi.
\end{eqnarray*}
\end{lemma}

\begin{proof}
A straightforward calculation gives that each Lebesgue integral on the right side of \eqref{eq:omegan} is differentiable with respect to the temperature $T$ under the integral sign. We thus obtain the result.
\end{proof}

We immediately have the following.

\begin{lemma}
The specific heat at constant volume corresponding to normal conductivity at the transition temperature $T_c$ is given by
\begin{eqnarray*}
C_V^N(T_c)
&=& 8T_c \int_{\varepsilon/(2T_c)}^{\hslash\omega_D/(2T_c)}
 \frac{N_0 \, \eta^2}{\, \cosh^2 \eta \,} \, d\eta
 + 4T_c \int_{-\mu/(2T_c)}^{-\hslash\omega_D/(2T_c)}
 \frac{N(2T_c \eta) \, \eta^2}{\, \cosh^2 \eta \,} \, d\eta \\
& & + 4T_c \int_{\hslash\omega_D/(2T_c)}^{\infty}
 \frac{N(2T_c\eta) \, \eta^2}{\, \cosh^2 \eta \,} \, d\eta.
\end{eqnarray*}
\end{lemma}

Since the gap $\Delta C_V$ in the specific heat at constant volume at $T=T_c$ is given by (see \cite[Proposition 2.5]{watanabe-five})
\[
\Delta C_V=-\frac{N_0}{\, 8\, T_c\,} \, \int_{\varepsilon/(2T_c)}^{\hslash\omega_D/(2T_c)}
 v(2\, T_c\, \eta)^2 g(\eta) \, d\eta \quad (>0),
\]
Theorem \ref{thm:dcv-over-cn} follows. 

\bigskip

\textit{Proof of Corollary \ref{crl:dcv-over-cn} } \    We then give a proof of Corollary \ref{crl:dcv-over-cn}. In many superconductors, the value $\hslash\omega_D/(2T_c)$ is very large, and hence the value $\mu/(2T_c)$ is also very large since $\mu>\hslash\omega_D$. So, if $\hslash\omega_D/(2\, T_c) \simeq \infty$ and $\varepsilon/(2\, T_c) \simeq 0$, then  the second and third terms of $J$ in Theorem \ref{thm:dcv-over-cn} are both very small since $\eta^2/\cosh^2 \eta \to 0$ as $\eta \to \pm \infty$. So
\[
J \simeq 2N_0 \, \int_{\varepsilon/(2\, T_c)}^{\hslash\omega_D/(2\, T_c)} \frac{\eta^2}{\, \cosh^2 \eta \,} \, d\eta.
\]
Moreover, in some superconductors, the value $U(x,\,\xi)$ is nearly equal to a constant at all $(x,\,\xi) \in [\varepsilon,\, \hslash\omega_D]^2$. So we set $U(x,\,\xi)=U_0$ at all $(x,\,\xi) \in [\varepsilon,\, \hslash\omega_D]^2$, where $U_0>0$ is a constant. Then the solution $u_0$ of Theorem \ref{thm:solution} does not depend on the energy $x$ and becomes a function of the temperature $T$ only. Accordingly, the function $v$ of Remark \ref{rmk:solution} becomes a constant $v_0>0$ since the function $v$ does not depend on the energy $x$. Hence $v(2T_c\,\eta)$ of Theorem \ref{thm:dcv-over-cn} becomes a constant, i.e., $v(2T_c\,\eta)=v_0$. Therefore Theorem \ref{thm:dcv-over-cn} implies
\[
\frac{\Delta C_V}{ \, C_V^N(T_c) \, } \simeq \frac{\, \displaystyle{ v_0^2 \, \int_{\varepsilon/(2\, T_c)}^{\hslash\omega_D/(2\, T_c)} \{ -g(\eta) \} \, d\eta } \,}{\, 64 \, T_c^2 \, \displaystyle{ \int_{\varepsilon/(2\, T_c)}^{\hslash\omega_D/(2\, T_c)} \frac{\eta^2}{\, \cosh^2 \eta \,} \, d\eta }  \,}.
\]
Note that (see \cite[Proposition 2.2]{watanabe-six})
\[
\frac{ \, v_0 \,}{\, 8 \, T_c \,}=\frac{\,  \displaystyle{ \tanh \frac{\, \hslash\omega_D \,}{2T_c}
 -\tanh \frac{\, \varepsilon \,}{\, 2T_c \,}  } \,}{\, \displaystyle{ \int_{\varepsilon/(2\, T_c)}^{\hslash\omega_D/(2\, T_c)} \{ -g(\eta) \} \, d\eta }  \,}.
\]
Here, $f'(T_c)$ and $\varepsilon$ in \cite[Proposition 2.2]{watanabe-six} are replaced by $-v_0$ and $\varepsilon/(2T_c)$, respectively. Then
\[
\frac{\Delta C_V}{ \, C_V^N(T_c) \, } \simeq
 \frac{\,  \displaystyle{ \left( \tanh \frac{\, \hslash\omega_D \,}{2T_c}
 -\tanh \frac{\, \varepsilon \,}{\, 2T_c \,} \right)^2  } \,}{\, 
 \displaystyle{ \int_{\varepsilon/(2\, T_c)}^{\hslash\omega_D/(2\, T_c)} \frac{\eta^2}{\, \cosh^2 \eta \,} \, d\eta } \; \displaystyle{ \int_{\varepsilon/(2\, T_c)}^{\hslash\omega_D/(2\, T_c)} \{ -g(\eta) \} \, d\eta }  \,}.
\]
Set $\hslash\omega_D/(2\, T_c) \simeq \infty$ and $\varepsilon/(2\, T_c) \simeq 0$, as mentioned above. Thus
\[
\frac{\Delta C_V}{ \, C_V^N(T_c) \, } \simeq
 \frac{\,  \pi^2      \,}{\, 
 \displaystyle{ 4 \, \int_0^{\infty} \frac{\eta^2}{\, \cosh^2 \eta \,} \, d\eta } \; \displaystyle{ \int_0^{\infty} \{ -g(\eta) \} \, d\eta }  \,},
\]
which does not depend on superconductors and is a universal constant. This proves Corollary \ref{crl:dcv-over-cn}.

\section{Proof of Theorem \ref{thm:cmag}}

We prove Theorem \ref{thm:cmag} in this section. Let us recall here that $\tau$ is very close to $T_c$.

\begin{lemma}\label{lm:psineg}
Let $\Psi(\cdot)$ be as in \eqref{eqn:superthermopotential}. Then $\Psi(T)<0$ at $T \in [\tau,\,T_c]$.
\end{lemma}

\begin{proof}
The equalities $\Psi(T_c)= (\partial \Psi/\partial T)(T_c)=0$ hold true (see \cite[Lemma 4.3]{watanabe-five}). Since $\Psi(\cdot) \in C^2[\tau,\,T_c]$ (see \cite[Lemma 4.5]{watanabe-five}), it follows that
\begin{eqnarray}\label{eq:taylor-psi}
\Psi(T)
&=& \Psi(T_c)+(T-T_c) \frac{\,\partial \Psi \,}{\partial T}(T_c)+\frac{\, (T-T_c)^2 \,}{2}
 \frac{\,\partial^2 \Psi \,}{\partial T^2}(c) \\
&=& \frac{\, (T-T_c)^2 \,}{2}
 \frac{\,\partial^2 \Psi \,}{\partial T^2}(c). \nonumber
\end{eqnarray}
Here, $c$ is between $T$ and $T_c$. By \cite[Lemma 4.5]{watanabe-five}, $(\partial^2 \Psi/\partial T^2)(T_c)<0$. Note that $(\partial^2 \Psi/\partial T^2)$ is continuous and that $T$ is very close to $T_c$. Hence the result follows.
\end{proof}

\begin{lemma}
Let $H_c(\cdot)$ be the critical magnetic field. Then $H_c(\cdot) \in C^1[\tau,\,T_c]$.
\end{lemma}

\begin{proof}
\textit{Step 1.} \hbox{} Since $\Psi(\cdot) \in C^2[\tau,\,T_c]$, Lemma \ref{lm:psineg} implies that $H_c(\cdot)=\sqrt{ -8\pi\Psi(\cdot) }$ is well-defined and continuous on $[\tau,\,T_c]$ and that $H_c(\cdot)$ is differentiable at $T \in [\tau,\,T_c)$. Here, its derivative is given by
\begin{equation}\label{eq:hdderivative}
\frac{\,\partial H_c \,}{\partial T}(T)=-4\pi \frac{\, (\partial \Psi/\partial T)(T) \,}{\,\sqrt{ -8\pi\Psi(T) }\,}.
\end{equation}
By $\Psi(T_c)=0$, the derivative \eqref{eq:hdderivative} is not defined at $T=T_c$. Hence we have only to show that it is differentiable at $T=T_c$. The equality $\Psi(T_c)=0$ implies $H_c(T_c)=\sqrt{ -8\pi\Psi(T_c) }=0$, and hence
\[
\frac{\, H_c(T_c)-H_c(T) \,}{T_c-T}=-\frac{\, \sqrt{ -8\pi\Psi(T) } \,}{T_c-T}=-2\sqrt{ -\pi \frac{\,\partial^2 \Psi \,}{\partial T^2}(c) }
\]
by \eqref{eq:taylor-psi}. Here, $c$ is between $T$ and $T_c$. Note that $(\partial^2 \Psi/\partial T^2)$ is continuous on $[\tau,\,T_c]$. Therefore
\[
\lim_{T \uparrow T_c} \frac{\, H_c(T_c)-H_c(T) \,}{T_c-T}=-2\sqrt{ -\pi \frac{\,\partial^2 \Psi \,}{\partial T^2}(T_c) },
\]
and hence $H_c(\cdot)$ is differentiable also at $T=T_c$, and
\[
\frac{\,\partial H_c \,}{\partial T}(T_c)=-2\sqrt{ -\pi \frac{\,\partial^2 \Psi \,}{\partial T^2}(T_c) }.
\]

\textit{Step 2.} \hbox{} We next show that the derivative $(\partial H_c/\partial T)$ is continuous on $[\tau,\,T_c]$. Since $\Psi(\cdot) \in C^2[\tau,\,T_c]$, it follows from \eqref{eq:hdderivative} that $(\partial H_c/\partial T)$ is continuous at $T \in [\tau,\,T_c)$. Hence we have only to show that the derivative is continuous at $T=T_c$. Since $\Psi(\cdot) \in C^2[\tau,\,T_c]$, it follows
\begin{eqnarray}\label{eq:taylor-d-psi}
\frac{\,\partial \Psi \,}{\partial T}(T)
&=& \frac{\,\partial \Psi \,}{\partial T}(T_c)+(T-T_c) \frac{\,\partial^2 \Psi \,}{\partial T^2}(c_1) \\
&=& (T-T_c) \frac{\,\partial^2 \Psi \,}{\partial T^2}(c_1), \nonumber
\end{eqnarray}
where $c_1$ is between $T$ and $T_c$. Combining \eqref{eq:hdderivative} with \eqref{eq:taylor-psi} and \eqref{eq:taylor-d-psi} yields
\begin{equation}\label{eq:hct-derivative}
\frac{\,\partial H_c \,}{\partial T}(T)=\sqrt{4\pi}
 \frac{(\partial^2 \Psi/\partial T^2)(c_1)}{\, \sqrt{ \, -(\partial^2 \Psi/\partial T^2)(c) \, } \,}
 \to -2\sqrt{ -\pi \frac{\,\partial^2 \Psi \,}{\partial T^2}(T_c) }=\frac{\,\partial H_c \,}{\partial T}(T_c)
 \quad \hbox{as} \quad T \uparrow T_c.
\end{equation}
So $(\partial H_c/\partial T)$ is continuous also at $T=T_c$.
\end{proof}

Let us recall here that  (see \cite[Lemma 4.5]{watanabe-five})
\[
\frac{\,\partial^2 \Psi \,}{\partial T^2}(T_c)=\frac{N_0}{\, 8\, T_c^2 \,} \, \int_{\varepsilon/(2T_c)}^{\hslash\omega_D/(2T_c)}
 v(2\, T_c\, \eta)^2 g(\eta) \, d\eta \quad (<0)
\]
and that $g(\eta)<0$ (see \eqref{eq:fng}).

\begin{lemma}
$H_c(T_c)=0$, $\displaystyle{ \frac{\,\partial H_c \,}{\,\partial T \,}(T)<0 }$ at $T \in [\tau,\, T_c]$, and
\[
\frac{\,\partial H_c \,}{\,\partial T \,}(T_c)=
 -\sqrt{ \, -\frac{\pi N_0}{\, 2 \, T_c^2 \,}
 \int_{\varepsilon/(2\, T_c)}^{\hslash\omega_D/(2\, T_c)} v(2T_c\,\eta)^2 \, g(\eta)\, d\eta \,} \quad (< 0).
\]
\end{lemma}

\begin{proof}
Due to the preceding lemma, we have only to show the inequality $\displaystyle{ \frac{\,\partial H_c \,}{\,\partial T \,}(T)<0 }$ at $T \in [\tau,\, T_c)$. By \eqref{eq:hct-derivative},
\[
\frac{\,\partial H_c \,}{\partial T}(T)=\sqrt{4\pi}
 \frac{(\partial^2 \Psi/\partial T^2)(c_1)}{\, \sqrt{ \, -(\partial^2 \Psi/\partial T^2)(c) \, } \,}, \quad
 T \in [\tau,\,T_c),
\]
where each of $c$ and $c_1$ is between $T$ and $T_c$. As mentioned before, the function $(\partial^2 \Psi/\partial T^2)$ is continuous on $[\tau,\,T_c]$, and $(\partial^2 \Psi/\partial T^2)(T_c)<0$. Since $\tau$ is very close to $T_c$, it follows that $\displaystyle{ \frac{\,\partial H_c \,}{\,\partial T \,}(T)<0 }$ at $T \in [\tau,\, T_c)$.
\end{proof}

\begin{lemma}
If $T \simeq T_c$ \hbox{} $(T \leq T_c)$, then
\[
H_c(T) \simeq  \left( 1-\frac{T}{\, T_c \,} \right) \sqrt{ -\frac{\, \pi N_0 \,}{2} 
 \int_{\varepsilon/(2\, T_c)}^{\hslash\omega_D/(2\, T_c)} v(2T_c\,\eta)^2 \, g(\eta)\, d\eta } \quad (\geq 0).
\]
\end{lemma}

\begin{proof}
By \eqref{eq:taylor-psi},
\[
H_c(T)=\sqrt{ \, -8\pi \Psi(T) \,}=(T_c-T) \sqrt{ \, -4\pi \frac{\,\partial^2 \Psi \,}{\partial T^2}(c) \,}
\simeq (T_c-T) \sqrt{ \, -4\pi \frac{\,\partial^2 \Psi \,}{\partial T^2}(T_c) \,}.
\]
The result thus follows.
\end{proof}

We next consider the critical magnetic field on the interval $[0,\, \tau_3]$.

\begin{lemma}
Let $\Psi(\cdot)$ be as in \eqref{eqn:superthermopotential}. Then $\Psi(T)<0$ on $[0,\, \tau_3]$.
\end{lemma}

\begin{proof}
By \eqref{eqn:superthermopotential},
\begin{eqnarray*}
-\Psi(T) &=& N_0 \int_{\varepsilon}^{\hslash\omega_D} \left\{
\frac{ u_0(T,\,\xi)^2 }{\, \sqrt{ \xi^2+u_0(T,\,\xi)^2 } \,} \left( 1-\tanh \frac{\, \sqrt{ \xi^2+u_0(T,\,\xi)^2 } \,}{2T} \right) \right.  \nonumber \\
& & \qquad \qquad +\frac{\, \left( \sqrt{ \xi^2+u_0(T,\,\xi)^2 }-\xi \right)^2\,}{\, \sqrt{ \xi^2+u_0(T,\,\xi)^2} \,}
 \left. -4T \ln \frac{  1+e^{-\xi/T}  }{\, 1+e^{ -\sqrt{\,\xi^2+u_0(T,\, \xi)^2\,}/T } \,}
 \right\} \, d\xi. \nonumber
\end{eqnarray*}
The sum of the first and second terms of the integrand is nonnegative, while the third term becomes
\[
\left|  -4T \ln \frac{  1+e^{-\xi/T}  }{\, 1+e^{ -\sqrt{\,\xi^2+u_0(T,\, \xi)^2\,}/T } \,} \right| \leq 
4T \, \ln 2 \leq 4 \tau_3 \, \ln 2.
\]
Since $\tau_3>0$ is small enough, the result follows.
\end{proof}

\begin{lemma}\label{lm:cpsi}
Let $\Psi(\cdot)$ be as in \eqref{eqn:superthermopotential}. Then $\displaystyle{ \Psi \in C[0,\, \tau_3]  }$.
\end{lemma}

\begin{proof}
A straightforward calculation gives that $\Psi(\cdot)$ is continuous at $T \in (0,\, \tau_3]$. So we have only to show its continuity at $T=0$. Then
\[
\sqrt{ \xi^2+u_0(T,\,\xi)^2 } \leq \sqrt{ \hslash^2\omega_D^2+\Delta_2(0)^2  },
\]
where the right side is integrable on $[\varepsilon,\, \hslash\omega_D]$. Moreover,
\[
\frac{u_0(T,\, \xi)^2}{\,\sqrt{ \xi^2+u_0(T,\, \xi)^2 } \,} \, \tanh \frac{\,\sqrt{\,\xi^2+u_0(T,\, \xi)^2\,}\,}{2T}
\leq u_0(T,\, \xi) \leq \Delta_2(0),
\]
where $\Delta_2(0)$ is also integrable on $[\varepsilon,\, \hslash\omega_D]$. As mentioned before,
\[
\left|  -4T \ln \frac{  1+e^{-\xi/T}  }{\, 1+e^{ -\sqrt{\,\xi^2+u_0(T,\, \xi)^2\,}/T } \,} \right| \leq 
4T \, \ln 2 \leq 4 \tau_3 \, \ln 2,
\]
where $4 \tau_3 \, \ln 2$ is again integrable on $[\varepsilon,\, \hslash\omega_D]$. Therefore the Lebesgue dominated convergence theorem gives 
\begin{eqnarray*}
\lim_{T \downarrow 0} \Psi(T) &=& -N_0 \lim_{T \downarrow 0} \int_{\varepsilon}^{\hslash\omega_D}
 \left\{
 \frac{ u_0(T,\,\xi)^2 }{\, \sqrt{ \xi^2+u_0(T,\,\xi)^2 } \,} \left( 1-\tanh \frac{\, \sqrt{ \xi^2+u_0(T,\,\xi)^2 } \,}{2T}
 \right) \right. \nonumber \\
& & \qquad \qquad +\left. \frac{\, \left( \sqrt{ \xi^2+u_0(T,\,\xi)^2 }-\xi \right)^2  \,}{\, \sqrt{ \xi^2+u_0(T,\,\xi)^2 } \,}
 -4T \ln \frac{  1+e^{-\xi/T}  }{\, 1+e^{ -\sqrt{\,\xi^2+u_0(T,\, \xi)^2\,}/T } \,}
 \right\} \, d\xi \nonumber \\
&=& -N_0 \int_{\varepsilon}^{\hslash\omega_D}
 \frac{\, \{ \, \sqrt{ \xi^2+u_0(0,\,\xi)^2 }-\xi \, \}^2  \,}{\, \sqrt{ \xi^2+u_0(0,\,\xi)^2 } \,}
 \, d\xi \nonumber \\
&=& \Psi(0).
\end{eqnarray*}
Thus $\Psi(\cdot)$ is continuous also at $T=0$.
\end{proof}

\begin{lemma}
Let $\Psi(\cdot)$ be as in \eqref{eqn:superthermopotential}. Then $\displaystyle{ \Psi \in C^1[0,\, \tau_3]  }$.
\end{lemma}

\begin{proof}
\textit{Step 1.} \  A straightforward calculation gives that $\Psi(\cdot)$ is differentiable at $T \in (0,\, \tau_3]$. So we need to show that $\Psi(\cdot)$ is differentiable also at $T=0$. Then
\[
\frac{ \, \Psi(T)-\Psi(0) \, }{T-0} = N_0 \int_{\varepsilon}^{\hslash\omega_D} 
 (  I+J ) \, d\xi, 
\]
where
\begin{eqnarray*}
I &=& \frac{2}{\, T \,} \left\{  \sqrt{ \xi^2+u_0(0,\,\xi)^2 }-\sqrt{ \xi^2+u_0(T,\,\xi)^2 }  \right\}, \\
J &=& \frac{1}{\, T \,} \left\{
 \frac{ u_0(T,\,\xi)^2 }{\, \sqrt{ \xi^2+u_0(T,\,\xi)^2 } \,} \tanh \frac{\, \sqrt{ \xi^2+u_0(T,\,\xi)^2 } \,}{2T}
   -\frac{ u_0(0,\,\xi)^2 }{\, \sqrt{ \xi^2+u_0(0,\,\xi)^2 } \,} \right\}.
\end{eqnarray*}
Since $u_0$ is partially differentiable with respect to $T$ on $[0,\, \tau_3] \times [\varepsilon,\, \hslash\omega_D]$, it follows that
\[
\left| \frac{ \, u_0(0,\,\xi)-u_0(T,\,\xi) \,  }{ T } \right| \leq \left| \frac{ \, \partial u_0 \,  }{ \partial T }(0,\, \xi) \right| +1.
\]
Here, $T>0$ is small enough. Therefore,
\begin{eqnarray*}
| I |
&=& 2 \frac{ u_0(0,\,\xi)+u_0(T,\,\xi) }{ \, \sqrt{ \xi^2+u_0(0,\,\xi)^2 }+\sqrt{ \xi^2+u_0(T,\,\xi)^2 } \, }
      \left| \frac{ \, u_0(0,\,\xi)-u_0(T,\,\xi) \,  }{ T } \right| \\
&\leq& 2 \frac{ \, \Delta_2(0) \, }{ \xi } 
 \left\{ \left| \frac{ \, \partial u_0 \,  }{ \partial T }(0,\, \xi) \right| +1 \right\},
\end{eqnarray*}
where the last term is integrable on $[\varepsilon,\, \hslash\omega_D]$. Hence the Lebesgue dominated convergence theorem gives
\[
\lim_{T \downarrow 0} N_0 \int_{\varepsilon}^{\hslash\omega_D} I \, d\xi=
 -N_0 \int_{\varepsilon}^{\hslash\omega_D}
 \frac{ 2\, u_0(0,\,\xi)  }{ \, \sqrt{ \xi^2+u_0(0,\,\xi)^2 } \, } \, \frac{ \, \partial u_0 \,  }{ \partial T }(0,\, \xi) 
 \, d\xi.
\]
A similar argument gives
\[
\lim_{T \downarrow 0} N_0 \int_{\varepsilon}^{\hslash\omega_D} J \, d\xi=
 N_0 \int_{\varepsilon}^{\hslash\omega_D} \left\{
 \frac{ 2\, u_0(0,\,\xi)  }{ \, \sqrt{ \xi^2+u_0(0,\,\xi)^2 } \, }
  -\frac{ u_0(0,\,\xi)^3  }{ \, \left( \xi^2+u_0(0,\,\xi)^2 \right)^{3/2} \, }
 \right\} \, \frac{ \, \partial u_0 \,  }{ \partial T }(0,\, \xi) \, d\xi.
\]
Thus $\Psi(\cdot)$ is differentiable at $T=0$, and
\begin{equation}\label{eq:psi-t-zero}
\frac{ \, \partial \Psi \, }{ \partial T }(0)=-N_0 \int_{\varepsilon}^{\hslash\omega_D}
   \frac{ u_0(0,\,\xi)^3  }{ \, \left( \xi^2+u_0(0,\,\xi)^2 \right)^{3/2} \, } \, 
   \frac{ \, \partial u_0 \,  }{ \partial T }(0,\, \xi) \, d\xi=0,
\end{equation}
since $\displaystyle{ \frac{ \, \partial u_0 \,  }{ \partial T }(0,\, \xi)=0  }$ at all $\xi \in [\varepsilon,\, \hslash\omega_D]$.

\textit{Step 2.} \  A straightforward calculation gives that $(\partial \Psi/\partial T)$ is continuous at $T \in (0,\, \tau_3]$. Here, $(\partial \Psi/\partial T)$ is given by \quad $( \, T \in (0,\, \tau_3] \, )$
\begin{eqnarray}\label{eq:psi-derivatine-t}
\frac{\, \partial \Psi \,}{ \partial T }(T) &=& N_0 \int_{\varepsilon}^{\hslash\omega_D} \left\{
 -\frac{ 2\, u_0(T,\,\xi)  }{ \, \sqrt{ \xi^2+u_0(T,\,\xi)^2 } \, } \, \frac{ \, \partial u_0 \,  }{ \partial T }(T,\, \xi)
 \left( 1-\tanh \frac{\, \sqrt{ \xi^2+u_0(T,\,\xi)^2 } \,}{2T}  \right)  \right. \\
& & \qquad \left. -\frac{ u_0(T,\,\xi)^3  }{ \, \left( \xi^2+u_0(T,\,\xi)^2 \right)^{3/2} \, } \, \frac{ \, \partial u_0 \,  }{ \partial T }(T,\, \xi) \, \tanh \frac{\, \sqrt{ \xi^2+u_0(T,\,\xi)^2 } \,}{2T} + K \right\} \, d\xi, \nonumber
\end{eqnarray}
where
\begin{eqnarray*}
K &=& \frac{ u_0(T,\,\xi)^2 }{\, 2T \, \left\{  \xi^2+u_0(T,\,\xi)^2 \right\} \,
 \cosh^2 \frac{\, \sqrt{ \xi^2+u_0(T,\,\xi)^2 } \,}{2T}  \, }
  \left( u_0(T,\,\xi) \frac{ \, \partial u_0 \,  }{ \partial T }(T,\, \xi)-\frac{ \, \xi^2+u_0(T,\,\xi)^2 \, }{T} \right) \\
& & +4 \ln \frac{  1+e^{-\xi/T}  }{\, 1+e^{ -\sqrt{\,\xi^2+u_0(T,\, \xi)^2\,}/T } \,}
      +4 \frac{ \xi }{\, T \left( 1+e^{\xi/T} \right)\, } \\
& & +4 \frac{ 1 }{\, 1+e^{\sqrt{\,\xi^2+u_0(T,\, \xi)^2\,}/T } \, } \left(
       \frac{ u_0(T,\,\xi)  }{ \, \sqrt{ \xi^2+u_0(T,\,\xi)^2 } \, } \, \frac{ \, \partial u_0 \,  }{ \partial T }(T,\, \xi)
     -\frac{ \, \sqrt{  \xi^2+u_0(T,\,\xi)^2 } \, }{T}     \right).
\end{eqnarray*}
So we have only to show that $(\partial \Psi/\partial T)$ is continuous at $T=0$. The first term of the integrand above becomes
\begin{eqnarray*}
& &\left| \frac{ 2\, u_0(T,\,\xi) }{\, \sqrt{ \xi^2+u_0(T,\,\xi)^2 } \, }\, \frac{\, \partial u_0 \,  }{ \partial T }(T,\, \xi)
 \left( 1-\tanh \frac{\, \sqrt{ \xi^2+u_0(T,\,\xi)^2 } \,}{2T}  \right)  \right| \\
&\leq&  2 \max_{(T,\,\xi) \in [0,\, \tau_3] \times [\varepsilon,\, \hslash\omega_D]}
 \left| \frac{ \, \partial u_0 \,  }{ \partial T }(T,\, \xi) \right| \, \frac{2T}{\, \sqrt{ \xi^2+u_0(T,\,\xi)^2 } \,} \\
&\leq& 2 \max_{(T,\,\xi) \in [0,\, \tau_3] \times [\varepsilon,\, \hslash\omega_D]}
 \left| \frac{ \, \partial u_0 \,  }{ \partial T }(T,\, \xi) \right| \,  \frac{\, 2\tau_3 \,}{\, \xi \,}.\,
\end{eqnarray*}
which does not depend on $T$ and is integrable on $[\varepsilon,\, \hslash\omega_D]$. We can deal with the rest of the integrand similarly. Hence the Lebesgue dominated convergence theorem gives
\[
\lim_{T \downarrow 0} \frac{\, \partial \Psi \,}{ \partial T }(T)
=-N_0 \int_{\varepsilon}^{\hslash\omega_D}
   \frac{ u_0(0,\,\xi)^3  }{ \, \left( \xi^2+u_0(0,\,\xi)^2 \right)^{3/2} \, } \, 
   \frac{ \, \partial u_0 \,  }{ \partial T }(0,\, \xi) \, d\xi
=\frac{ \, \partial \Psi \, }{ \partial T }(0) \; (=0).
\]
Therefore, $(\partial \Psi/\partial T)$ is continuous at $T=0$, and hence at all $T \in [0,\, \tau_3]$.
\end{proof}

On the basis of the lemmas above, we now turn to the critical magnetic field $H_c(\cdot)$ on the interval $[0,\, \tau_3]$.

\begin{lemma}
$H_c(\cdot) \in C^1[0,\, \tau_3]$, and 
\[
H_c(0) = \sqrt{ \;  8\pi N_0 \int_{\varepsilon}^{\hslash\omega_D}
 \frac{\, \{ \, \sqrt{ \xi^2+u_0(0,\,\xi)^2 }-\xi \, \}^2  \,}{\, \sqrt{ \xi^2+u_0(0,\,\xi)^2 } \,}
 \, d\xi  \; }.
\]
\end{lemma}

\begin{proof}
The lemmas above immediately give that $H_c(\cdot) \in C^1[0,\, \tau_3]$. Since $H_c(0)=\sqrt{ -8 \pi \Psi(0)  }$, the rest follows from the argument in the proof of Lemma \ref{lm:cpsi}.
\end{proof}

\begin{lemma}
$(\partial H_c/\partial T)(T)<0$ at $T \in (0,\,\tau_3]$, and
\[
\frac{\,\partial H_c \,}{\,\partial T \,}(0)=0.
\]
\end{lemma}

\begin{proof}
Note that $(\partial u/\partial T)(T,\, \xi)<0$ at $T \in (0,\, \tau_3]$ and that $\tau_3>0$ is small enough. The fact that $| K |$ of \eqref{eq:psi-derivatine-t} is small enough for $T \in [0,\, \tau_3]$ gives that $(\partial \Psi/\partial T)(T)>0$ at $T \in (0,\, \tau_3]$. Therefore, by \eqref{eq:hdderivative},
\[
\frac{\,\partial H_c \,}{\partial T}(T)=-4\pi \frac{\, (\partial \Psi/\partial T)(T) \,}{\,\sqrt{ -8\pi\Psi(T) }\,}
<0, \qquad T \in (0,\, \tau_3].
\]
The result $(\partial H_c/\partial T)(0)=0$ follows from \eqref{eq:psi-t-zero}.
\end{proof}

Our proof of Theorem \ref{thm:cmag} is complete.


\begin{figure}[htbp]
\includegraphics[width=16cm]{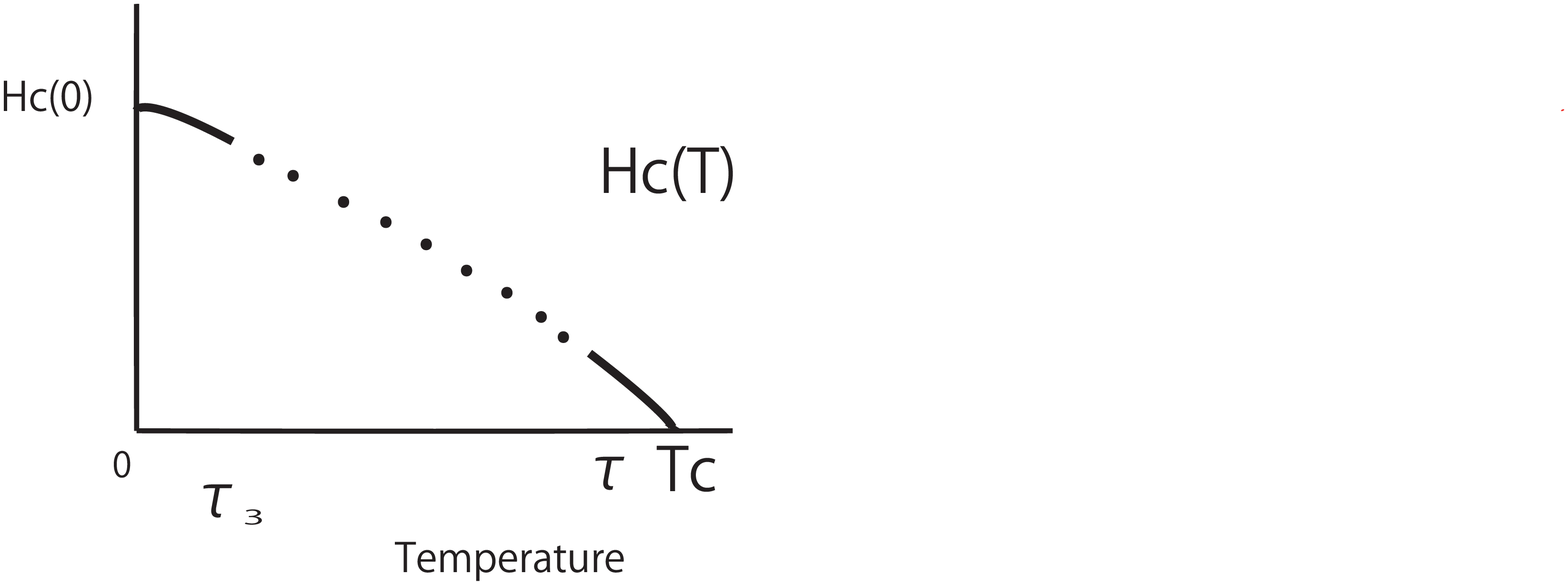}
\vspace{-3cm}
\caption{\textsf{The behavior of $H_c(T)$}}
\end{figure}

\noindent \textbf{Additional Information}

The author declares no competing interests.


\begin{thebibliography}{99}

\bibitem{bls} Bach, V., Lieb, E.~H. \&  Solovej, J.~P.
Generalized Hartree-Fock theory and the Hubbard model.
\textit{Nat. Rev. Genet J. Stat. Phys.} \textbf{76}, 3--89 (1994).

\bibitem{bcs} Bardeen, J., Cooper, L.~N. \& Schrieffer, J.~R.
Theory of superconductivity.
\textit{Phys. Rev.} \textbf {108}, 1175--1204 (1957).

\bibitem{billardfano} Billard, P. \& Fano, G.
An existence proof for the gap equation in the superconductivity theory.
\textit{Commun. Math. Phys.} \textbf{10}, 274--279 (1968).

\bibitem{bogoliubov}  Bogoliubov, N.~N.
A new method in the theory of superconductivity I.
\textit{Soviet Phys. JETP} \textbf {34}, 41--46 (1958).

\bibitem{chen} Chen, T., Fr$\ddot{\hbox{o}}$hlich, J. \& Seifert, M.
Renormalization Group Methods: Landau-Fermi Liquid and BCS Superconductor.
Proc. of the 1994 Les Houches Summer School. arXiv:cond-mat/9508063.

\bibitem{deugeihailoss} Deuchert, A., Geisinger, A., Hainzl, C. \&  Loss, M.
 Persistence of translational symmetry in the BCS model with radial pair interaction.
\textit{Ann. Henri. Poincar\'e} \textbf {19}, 1507--1527 (2018).

\bibitem{fhns}  Frank, R.~L., Hainzl, C., Naboko, S. \& Seiringer, R.
The critical temperature for the BCS equation at weak coupling.
\textit{J. Geom. Anal.} \textbf{17}, 559--568 (2007).

\bibitem{fhss} Frank, R.~L., Hainzl, C., Seiringer, R. \& Solovej, J.~P.
The external field dependence of the BCS critical temperature.
\textit{Commun. Math. Phys.} \textbf {342}, 189--216 (2016).

\bibitem{freijihaizlseiringer} Freiji, A., Hainzl, C. \& Seiringer, R.
The gap equation for spin-polarized fermions.
\textit{J. Math. Phys.} \textbf {53}, 012101 (2012).

\bibitem{hhss}  Hainzl, C., Hamza, E., Seiringer, R. \& Solovej, J.~P.
The BCS functional for general pair interactions.
\textit{Commun. Math. Phys.} \textbf {281}, 349--367 (2008).

\bibitem{hainzlloss} Hainzl, C. \& Loss, M.
General pairing mechanisms in the BCS-theory of superconductivity.
\textit{Eur. Phys. J. B}, 90:82 (2017).

\bibitem{haizlseiringer} Hainzl, C. \& Seiringer, R.
Critical temperature and energy gap for the BCS equation.
\textit{Phys. Rev.} \textbf {B 77}, 184517 (2008).

\bibitem{haizlseiringer2}  Hainzl, C. \& Seiringer, R. 
The BCS critical temperature for potentials with negative scattering length.
\textit{Lett. Math. Phys.} \textbf {84}, 99--107 (2008).

\bibitem{haizlseiringer3} Hainzl, C. \& Seiringer, R.
The Bardeen-Cooper-Schrieffer functional of superconductivity and its mathematical properties.
\textit{J. Math. Phys.} \textbf {57}, 021101 (2016).

\bibitem{kuzemsky}  Kuzemsky, A.~L.
Bogoliubov's vision: quasiaverages and broken symmetry to quantum protectorate and emergence.
\textit{Internat. J. Mod. Phys.} \textbf{B 24}, 835--935 (2010).

\bibitem{kuzemsky2}  Kuzemsky, A.~L.
Variational principle of Bogoliubov and generalized mean fields in many-particle interacting systems.
\textit{Internat. J. Mod. Phys.} \textbf{B  29}, 1530010 (63 pages)  (2015).

\bibitem{kuzemsky3} Kuzemsky, A.~L.
Statistical Mechanics and the Physics of Many-Particle Model Systems.
(World Scientific Publishing Co, 2017).

\bibitem{maskawa-nakajima-one} Maskawa, T. \& Nakajima, H.
Spontaneous breaking of chiral symmetry in a vector-gluon model.
\textit{Prog. Theor. Phys.} \textbf{52}, 1326--1354 (1974).

\bibitem{maskawa-nakajima-two} Maskawa, T. \& Nakajima, H. 
Spontaneous breaking of chiral symmetry in a vector-gluon model II.
\textit{Prog. Theor. Phys.} \textbf{54}, 860--877  (1975).

\bibitem{niwa} Niwa, M. 
Fundamentals of Superconductivity.
(Tokyo Denki University Press, 2002).

\bibitem{odeh} Odeh, F.
An existence theorem for the BCS integral equation.
\textit{IBM J. Res. Develop.} \textbf {8}, 187--188 (1964).

\bibitem{vansevenant} Vansevenant, A.
The gap equation in the superconductivity theory.
\textit{Physica} \textbf{17D}, 339--344  (1985).

\bibitem{watanabe-one} Watanabe, S.
The solution to the BCS gap equation and the second-order phase transition in superconductivity.
\textit{J. Math. Anal. Appl.} \textbf{383}, 353--364  (2011).

\bibitem{watanabe-two} Watanabe, S.
Addendum to `The solution to the BCS gap equation and the second-order phase transition in superconductivity'.
\textit{J. Math. Anal. Appl.} \textbf{405}, 742--745  (2013).

\bibitem{watanabe-three} Watanabe, S.
An operator-theoretical treatment of the Maskawa-Nakajima equation in the massless abelian gluon model.
\textit{J. Math. Anal. Appl.} \textbf{418}, 874--883 (2014).

\bibitem{watanabe-four} Watanabe, S. \& Kuriyama, K.
Smoothness and monotone decreasingness of the solution to the BCS-Bogoliubov gap equation for superconductivity.
\textit{J. Basic and Applied Sciences} \textbf{13}, 17--25  (2017).

\bibitem{watanabe-five} Watanabe, S. 
An operator-theoretical proof for the second-order phase transition in the BCS-Bogoliubov model of
superconductivity. 
\textit{Kyushu J. Math.} \textbf{74}, 177-196  (2020).

\bibitem{watanabe-six} Watanabe, S. 
A mathematical proof that the transition to a superconducting state is a second-order phase transition.
arXiv:0808.3438v1.

\bibitem{ziman} Ziman, J.~M. 
Principles of the Theory of Solids.
(Cambridge University Press, 1972).

\end{thebibliography}
\end{document}